\documentclass{article}
\usepackage[utf8]{inputenc}

\usepackage[T1]{fontenc}
\usepackage{soul, comment} 
\usepackage{phfqit}
\usepackage{proba} 
\usepackage[sort,nocompress]{cite}
\usepackage{amssymb}
\usepackage{amsmath}
\usepackage{amsthm}
\usepackage{amsfonts}
\usepackage{mathtools}
\usepackage{xspace}
\usepackage{bm}
\usepackage{nicefrac}
\usepackage{commath}
\usepackage{bbm}
\usepackage{boxedminipage}
\usepackage{xparse}
\usepackage{xcolor}
\usepackage{float}
\usepackage{multirow}
\usepackage{makecell}
\usepackage{graphicx}
\usepackage{caption}
\usepackage{subcaption}
\usepackage{ifthen}
\usepackage{thm-restate}
\usepackage{algorithm}
\usepackage[noend]{algpseudocode}
\usepackage{colortbl} 
\usepackage{tikz}
\usetikzlibrary{arrows.meta,positioning,calc,fit,backgrounds}
\tikzset{vtx/.style={circle,draw,fill=white,inner sep=2.4pt,font=\small},
 box/.style={draw,rounded corners=2pt,fill=blue!3,inner sep=5pt,font=\small},
 arr/.style={-{Stealth[length=2mm]},thick},
 aux/.style={-{Stealth[length=2mm]},thick,dashed,draw=warm}}

\definecolor{inkblue}{RGB}{28,77,117}
\definecolor{warm}{RGB}{166,86,28}

\usepackage{fancyhdr}   
\usepackage{hyperref}
    \hypersetup{ colorlinks=true, linkcolor=blue, citecolor=magenta, urlcolor=blue,}
\usepackage{fullpage}
\usepackage[utf8]{inputenc}
\usepackage{environ}
\usepackage{mdframed}
\usepackage{cleveref}
\usepackage[short]{optidef} 
\usepackage{tikz}
\usetikzlibrary{shapes}
\usetikzlibrary{positioning}
\usetikzlibrary{fit}
\usetikzlibrary{graphs,graphs.standard}
\usepackage{enumitem}

\newtheorem{proposition}{Proposition}[section]
\newtheorem{lemma}{Lemma}[section]

\newtheorem{theorem}{Theorem}[section]

\newtheorem{corollary}{Corollary}[section]

\newtheorem{definition}{Definition}[section]

\usepackage{enumitem}

\NewEnviron{problem}[1]{%
	\begin{center}\fbox{\parbox{6in}{%
				{\centering\scshape #1\par}%
				\parskip=1ex
				\everypar{\hangindent=1em}%
				\BODY
}}\end{center}}

\newcommand{\opt}{\operatorname{OPT}}

\newcommand{\compact}{\mathcal{N}_{\mathrm{compact}}}
\newcommand{\augment}{\mathcal{N}^{\mathrm{aug}}_f}
\newcommand{\augflow}{f^{\mathrm{aug}}}
\newcommand{\intaugflow}{f^{\mathrm{aug}}_{\mathrm{int}}}

\title{Canonical Cut Hierarchy in Almost-Linear Time via Reduction to Max-Flow}
\author{
Karthekeyan Chandrasekaran\thanks{University of Illinois, Urbana-Champaign, USA, Email: \{karthe, harshul4\}@illinois.edu. Supported in part by NSF grant CCF-2402667. }
\and Harshul Sagar\footnotemark[1]
}

\date{}
\begin{document}
\maketitle
\begin{abstract}
    The canonical cut hierarchy recursively partitions a graph along Tutte-Nash-Williams partitions, choosing the partition with the most number of parts at each step. We give a randomized almost linear-time algorithm for computing this hierarchy in undirected graphs with polynomially bounded capacities via reduction to max-flow. 
    This yields randomized almost linear-time algorithms for computing partition connectivity, strength, both arboricity and fractional arboricity, both integral and fractional spanning tree packing numbers, and implicit representations of the principal partition sequence and a maximum fractional spanning tree packing. 
\end{abstract}

\tableofcontents
\newpage
\pagenumbering{arabic}
\setcounter{page}{1}

\section{Introduction}\label{sec:intro}
Global minimum cut and Gomory-Hu trees are central to graph algorithms and combinatorial optimization. They capture fundamental connectivity information: global minimum cut measures the graph's edge connectivity while a Gomory-Hu tree compactly represents all-pairs minimum cuts \cite{GH61}. The pursuit of efficient algorithms to compute them has produced a rich toolkit of techniques. Karger \cite{Kar00} gave a randomized near-linear time algorithm for global min-cut which has subsequently been improved \cite{GMW24, MN20, GMW21, HLRW24, LP25}. 
For Gomory-Hu trees, an exciting recent sequence of breakthroughs culminated in almost linear-time algorithms \cite{AKT21, AKLPST22, Zha21, ALPS23, AKLPGSYY25}. The development of almost linear-time maximum flow algorithms for polynomially bounded integer capacities \cite{CKLPPS25, BCKLPPSS23} has advanced a fundamentally different approach for both problems---here, the problem is reduced to a small number of maximum-flow computations on graphs of small aggregate size, providing a flow-based route to fast algorithms \cite{LP25, ALPS23, AKLPGSYY25}. 

In this work, we design a fast flow-based algorithm for another fundamental graph structure, the \emph{canonical cut hierarchy}. This hierarchy is obtained by recursively decomposing the graph along Tutte-Nash-Williams partitions---we will define it formally shortly. 
An equivalent decomposition was studied by Trubin under the name HS-decomposition \cite{Tru93}, and the underlying decomposition also appears in Thorup's work on ideal tree packing \cite{Tho07,Tho08}.
The hierarchy is closely connected to the classical
principal partition sequence \cite{CQX20,HQC23}.
We follow the canonical formulation studied by Cen,
Fleischmann, Li, Li, and Panigrahi \cite{CFLLP25}, by choosing the finest Tutte--Nash--Williams
partition in each recursive step. 
The hierarchy captures strength, partition connectivity, both arboricity and fractional arboricity, both integral and fractional spanning tree packing numbers, and implicitly represents the principal partition sequence as well as a maximum fractional packing of spanning trees. Computing it efficiently, therefore, provides a common structural tool for several graph optimization problems. 

We now set up the background needed to define the canonical cut hierarchy of a graph. Let $G=(V, E, c: E\rightarrow \Z_+)$ be an undirected graph with $n$ vertices, $m$ edges and positive integer capacities. For a partition $\Pi$ of the vertex set into non-empty parts, let $\delta(\Pi)$ denote the crossing edges and $c(\delta(\Pi)):=\sum_{e\in \delta(\Pi)}c_e$. The \emph{strength} of $G$ is 
\begin{equation}\label{eq:strength}
    \beta(G):=\min_{\Pi: |\Pi|\ge 2}\frac{c(\delta(\Pi))}{|\Pi|-1}. 
\end{equation}
A minimizing partition for $\beta(G)$ is a \emph{Tutte-Nash-Williams partition}. 
The \emph{partition connectivity} of $G$ is $\lfloor \beta(G)\rfloor$. 
The \emph{tree packing number} of $G$ is the maximum number of spanning trees, counted with multiplicity, such that each edge $e$ occurs in at most $c_e$ trees. 
The \emph{fractional tree packing number} of $G$ is the maximum sum of weight of all spanning trees in an assignment of non-negative weights to spanning trees such that the total weight of the spanning trees using each edge $e$ is at most $c_e$. 
Tutte \cite{Tut61} and Nash-Williams \cite{NW61} showed that the tree packing number is equal to the partition connectivity, and it implies that the fractional tree packing number is equal to the strength. 

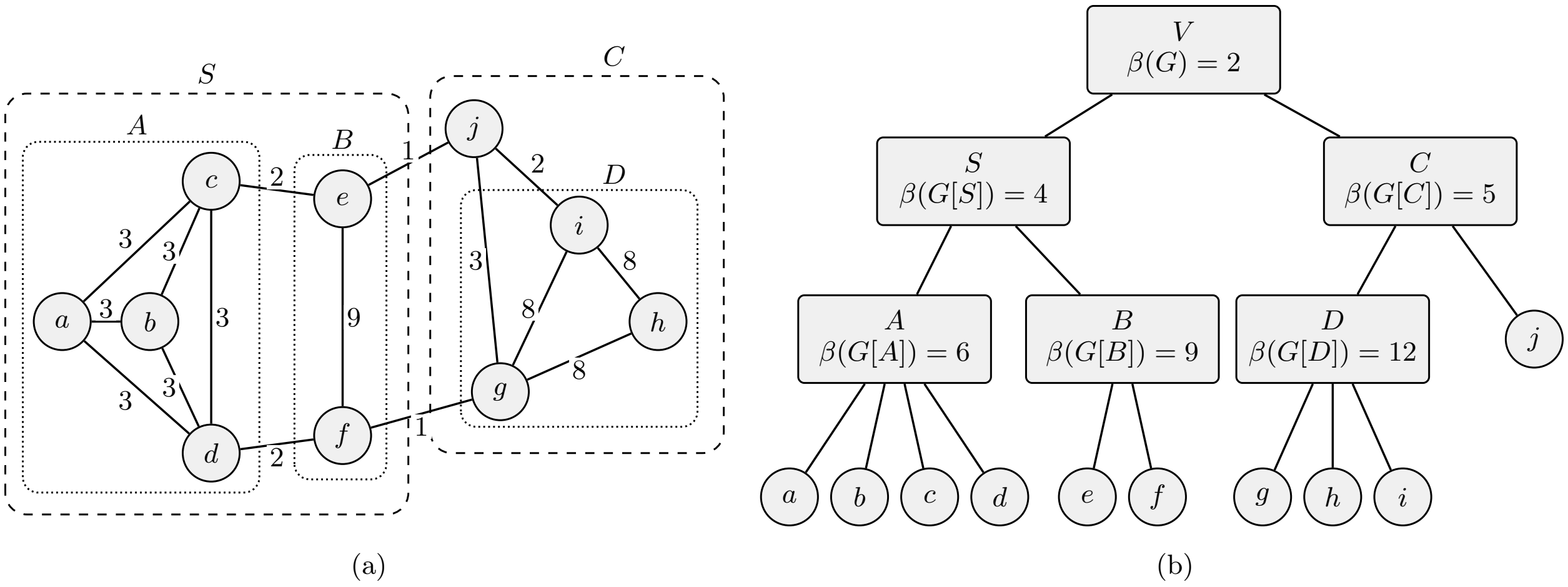
\begin{figure}[ht]
\centering
\resizebox{\linewidth}{!}{%
\begin{tikzpicture}[
  x=1cm,y=1cm,font=\small,
  vertex/.style={
    circle,draw=black,fill=gray!12,text=black,
    minimum size=6.5mm,inner sep=0pt,line width=0.6pt
  },
  hierarchy/.style={
    rectangle,rounded corners=2pt,draw=black,
    fill=gray!12,text=black,align=center,
    minimum width=22mm,minimum height=10mm,
    inner sep=2pt,line width=0.6pt
  },
  edge/.style={draw=black,line width=0.7pt},
  weight/.style={fill=white,inner sep=1pt,font=\small},
  part/.style={
    draw=black,densely dotted,rounded corners=5pt,
    line width=0.6pt
  },
  group/.style={
    draw=black,dashed,rounded corners=7pt,
    line width=0.6pt
  }
]

\draw[part] (0.05,1.05) rectangle (2.75,5.05);
\draw[part] (3.15,1.2) rectangle (4.2,4.9);
\draw[part] (5.05,1.8) rectangle (7.75,4.5);

\draw[group] (-0.15,0.8) rectangle (4.45,5.6);
\draw[group] (4.7,1.5) rectangle (8.05,5.8);

\node at (1.35,5.24) {$A$};
\node at (3.7,5.08) {$B$};
\node at (6.8,4.68) {$D$};
\node at (2.15,5.83) {$S$};
\node at (6.8,6.03) {$C$};

\node[vertex] (ga) at (0.5,3) {$a$};
\node[vertex] (gb) at (1.5,3) {$b$};
\node[vertex] (gc) at (2.2,4.6) {$c$};
\node[vertex] (gd) at (2.2,1.5) {$d$};
\node[vertex] (ge) at (3.7,4.4) {$e$};
\node[vertex] (gf) at (3.7,1.7) {$f$};
\node[vertex] (gg) at (5.5,2.2) {$g$};
\node[vertex] (gh) at (7.3,3) {$h$};
\node[vertex] (gi) at (6.4,4.1) {$i$};
\node[vertex] (gj) at (5.2,5.2) {$j$};

\draw[edge] (ga) -- node[weight,above] {$3$} (gb);
\draw[edge] (ga) -- node[weight,above left] {$3$} (gc);
\draw[edge] (ga) -- node[weight,below left] {$3$} (gd);
\draw[edge] (gb) -- node[weight,left] {$3$} (gc);
\draw[edge] (gb) -- node[weight,left] {$3$} (gd);
\draw[edge] (gc) -- node[weight,right] {$3$} (gd);

\draw[edge] (ge) -- node[weight,right] {$9$} (gf);

\draw[edge] (gg) -- node[weight,below] {$8$} (gh);
\draw[edge] (gh) -- node[weight,above right] {$8$} (gi);
\draw[edge] (gg) -- node[weight,left] {$8$} (gi);

\draw[edge] (gg) -- node[weight,left] {$3$} (gj);
\draw[edge] (gi) -- node[weight,above right] {$2$} (gj);

\draw[edge] (gc) -- node[weight,above] {$2$} (ge);
\draw[edge] (gd) -- node[weight,below] {$2$} (gf);

\draw[edge] (gf) -- node[weight,below] {$1$} (gg);
\draw[edge] (ge) -- node[weight,above] {$1$} (gj);

\node at (4,0.2) {(a)};

\node[hierarchy] (hV) at (13.3,6.1)
  {$V$\\[-1pt]$\beta(G)=2$};

\node[hierarchy] (hS) at (10.9,4.6)
  {$S$\\[-1pt]$\beta(G[S])=4$};

\node[hierarchy] (hC) at (16,4.6)
  {$C$\\[-1pt]$\beta(G[C])=5$};

\node[hierarchy] (hA) at (10,2.8)
  {$A$\\[-1pt]$\beta(G[A])=6$};

\node[hierarchy] (hB) at (12.6,2.8)
  {$B$\\[-1pt]$\beta(G[B])=9$};

\node[hierarchy] (hD) at (15,2.8)
  {$D$\\[-1pt]$\beta(G[D])=12$};

\node[vertex] (ha) at (8.8,1) {$a$};
\node[vertex] (hb) at (9.6,1) {$b$};
\node[vertex] (hc) at (10.4,1) {$c$};
\node[vertex] (hd) at (11.2,1) {$d$};
\node[vertex] (he) at (12.2,1) {$e$};
\node[vertex] (hf) at (13,1) {$f$};
\node[vertex] (hg) at (14.2,1) {$g$};
\node[vertex] (hh) at (15,1) {$h$};
\node[vertex] (hi) at (15.8,1) {$i$};
\node[vertex] (hj) at (17.3,2.8) {$j$};

\draw[edge] (hV) -- (hS);
\draw[edge] (hV) -- (hC);

\draw[edge] (hS) -- (hA);
\draw[edge] (hS) -- (hB);

\draw[edge] (hC) -- (hD);
\draw[edge] (hC) -- (hj);

\draw[edge] (hA) -- (ha);
\draw[edge] (hA) -- (hb);
\draw[edge] (hA) -- (hc);
\draw[edge] (hA) -- (hd);

\draw[edge] (hB) -- (he);
\draw[edge] (hB) -- (hf);

\draw[edge] (hD) -- (hg);
\draw[edge] (hD) -- (hh);
\draw[edge] (hD) -- (hi);

\node at (13.2,0.2) {(b)};

\end{tikzpicture}%
}
\caption{
A graph and its canonical cut hierarchy.
(a) Edge labels are capacities.
The dotted outlines mark $A=\{a,b,c,d\}$, $B=\{e,f\}$,
and $D=\{g,h,i\}$.
The dashed outlines mark $S=A\cup B$ and $C=D\cup\{j\}$.
(b) Each internal node names its represented set $X$ and
displays its strength $\beta(G[X])$.
Its children form the finest Tutte--Nash--Williams
partition of $G[X]$.
The root splits into $S$ and $C$.
The set $S$ splits into $A$ and $B$, whereas $C$ splits
into $D$ and the singleton $\{j\}$.
The sets $A$, $B$, and $D$ then split into singletons,
giving four, two, and three children, respectively.
Each represented set is specified implicitly by its
descendant leaves, so only child pointers and strength
labels need to be stored.
}
\label{fig:canonical-cut-hierarchy}
\end{figure}

The \emph{canonical cut hierarchy} is a rooted tree whose root represents $V$ and whose leaves represent the vertices of $G$. Each internal node represents the set $S\subseteq V$ of its descendant leaves and is labeled by the strength of $G[S]$, i.e., $\beta(G[S])$. The children of an internal node representing $S$ represent the parts of the unique finest Tutte-Nash-Williams partition of $G[S]$. 
Here, finest means that the partition has the largest number of parts among all Tutte-Nash-Williams partitions. This choice makes the hierarchy canonical \cite{CFLLP25}. Every internal node has at least two children, so there are at most $2n-1$ nodes in the tree. 
Thus, the canonical cut hierarchy can be stored in $O(n)$-space: each node stores its children, and for an internal node $S$, its strength $\beta(G[S])$. 
In this representation, the vertices represented by a node are its descendant leaves. See Figure~\ref{fig:canonical-cut-hierarchy} for an example. 

In addition to strength and partition connectivity, the hierarchy encodes arboricity and fractional arboricity. Its largest internal-node label is the \emph{fractional arboricity}, namely $\max_{S\subseteq V: |S|\ge 2}c(E[S])/(|S|-1)$, where $E[S]$ contains the edges within $S$. The ceiling of fractional arboricity is known as \emph{arboricity}. Nash-Williams \cite{NW64} showed that arboricity is equal to the minimum number of forests whose union contains each edge $e$ exactly $c_e$ times. 

An equivalent description connects the canonical cut hierarchy to the classical \emph{principal partition sequence}. For a parameter $\lambda \ge 0$, we consider the Lagrangian problem associated with strength, namely
\begin{align}
\Phi_{G, \lambda}(\Pi)&:=c(\delta(\Pi))-\lambda(|\Pi|-1) \text{ and } \notag\\
\opt_G(\lambda)&:=\min_{\Pi}\Phi_{G, \lambda}(\Pi). \label{eq:lagrangian}
\end{align}
We denote $\opt_G(\lambda)$ as the Lagrangian problem on $G$ for $\lambda$. 
We note that unlike strength, the Lagrangian for $\opt_G(\lambda)$ allows the partition $\Pi$ to have only one part. 
For a fixed parameter $\lambda$, we say that a partition is a finest optimum for $\opt_G(\lambda)$ if it is a minimizer for $\opt_G(\lambda)$ and has the largest number of parts among all such minimizers. 
For each $\lambda$, there is a unique finest optimum partition for $\opt_G(\lambda)$. The sequence $\Pi_1, \Pi_2, \Pi_3, \ldots,$ of unique finest optimum partitions for $\opt_G(\lambda)$  obtained as we vary $\lambda$ from $0$ to $\infty$ is the \emph{principal partition sequence}. It is known that the sequence has at most $n$ partitions and has a nested structure, namely parts of $\Pi_{i+1}$ are obtained by splitting some of the parts of $\Pi_i$. The $\lambda$ values where the optimum value changes are known as \emph{breakpoints}. The canonical cut hierarchy is a rooted tree representation of the principal partition sequence with nodes representing parts of the partition and node labels being breakpoints. The principal partition sequence has a long history in graph, matroid, and submodular optimization \cite{Nar91, Fuj09, Nar97}. Their applications include network reinforcement \cite{Cun85}, rigidity \cite{PN92}, minimum $k$-cut approximation for all positive integers $k$ \cite{NRP96, RS08, CQX20, Qua22}, and submodular $k$-partition \cite{CW24, BCKS26}. 
Together with the capacities, the hierarchy determines
Thorup's ideal edge loads \cite{Tho07,Tho08}. 
After scaling by the strength of the graph, these give
the edge loads of a maximum fractional spanning tree
packing---see \cite{CQX20} for details.
Thus, the hierarchy implicitly represents a maximum fractional spanning tree packing without listing its trees.
For these reasons, fast computation of the sequence is a well-investigated problem in the literature \cite{Nar91, Nar97, PN92, Kol10, Cun85, Gus91, CC94, Gab98, Bar92, Qua22, CFLLP25}. 

The fast algorithms for global minimum cut and Gomory-Hu trees lead to a natural question: can strength be computed in almost linear-time and can the entire canonical cut hierarchy be computed just as fast? The recursive definition uses $O(n)$ computations of strength and can repeatedly process large subgraphs. Yet, we answer both questions affirmatively: for polynomially bounded positive integer capacities, we show that strength and the canonical cut hierarchy can be computed in randomized almost linear-time. As a consequence, we can compute partition connectivity, both arboricity and fractional arboricity, both integral and fractional tree packing numbers, and implicit representations of the principal partition sequence and a maximum fractional tree packing in randomized almost linear time.

\subsection{Previous Work}
Approximate strength can be computed in near-linear time through approximate fractional tree packing. Approximate fractional tree packing is a key ingredient in Karger's global minimum cut algorithm \cite{Kar00}. Chekuri and Quanrud \cite{CQ17} give a deterministic $\tilde O(m/\varepsilon^2)$-time algorithm for a $(1-\varepsilon)$-approximate maximum fractional spanning-tree packing, and hence for approximating strength. Our focus is exact computation of strength and the entire canonical cut hierarchy. Throughout, $\widetilde{O}$ suppresses polylogarithmic factors and almost-linear time means $(m+n)^{1+o(1)}$. 

\paragraph{Strongly polynomial-time algorithms.}
The equivalence with the principal partition sequence connects the canonical cut hierarchy problem to classical results. 
Narayanan \cite{Nar91} showed that principal partition sequence of a graph can be computed using $O(n^2)$ maximum-flow computations. Patkar and Narayanan \cite{PN92} compute a structure that is richer than the principal partition sequence structure in $O(mn^2 \log^2 n)$  time. Kolmogorov \cite{Kol10} computes the principal partition sequence using $n$ \emph{parameteric maximum-flow} computations, yielding an $O(n^2m\log{n})$ time. 
For computing exact strength, Cunningham \cite{Cun85} initiated an algorithmic approach based on polymatroids and network flows. Gusfield \cite{Gus91} and Cheng and Cunningham\cite{CC94} used parametric-flow techniques to compute strength in $O(n^3m)$ and $O(n^4)$ time, respectively. Gabow \cite{Gab98} gave an $O(n^2 m\log{n})$ time algorithm for computing strength using parameteric-cut methods. 

\paragraph{Weakly polynomial-time algorithms.}
For a fixed $\lambda$, Barahona \cite{Bar92} solved the Lagrangian problem \eqref{eq:lagrangian} using at most $n$ maximum-flow computations. We observe that $\opt_G(\lambda)<0$ if and only if $\beta(G)<\lambda$. Combining this with binary search and modern maximum-flow algorithms gives an $nm^{1+o(1)}$-time algorithm to compute strength when capacities are polynomially bounded integers. Cen, Fleischmann, Li, Li, and Panigrahi \cite{CFLLP25} showed that the canonical cut hierarchy can be computed in $nm^{1+o(1)}$ time for polynomially bounded capacities. For comparison, our algorithm computes the canonical cut hierarchy in $m^{1+o(1)}$ time for polynomially bounded capacities.

\subsection{Our Results}
We recall the $O(n)$-sized representation for the canonical cut hierarchy: Each node stores its children, and for an internal node $S$, its strength $\beta(G[S])$; the vertices represented by a node are its descendant leaves. By canonical cut hierarchy, we mean such a representation. Our main result is a randomized almost linear-time algorithm to compute the canonical cut hierarchy of a given graph with polynomially bounded capacities.

\begin{theorem}\label{thm:cch}
    Let $G=(V, E, c: E\rightarrow \Z_+)$ be an undirected graph with $n$ vertices and $m$ edges. Let $W:=\sum_{e\in E}c_e$ and $L:=1+\lceil\log_2{((W+1)(n-1)^2)} \rceil$. For every $\eta\in (0, 1)$, there is a randomized algorithm that returns the canonical cut hierarchy of $G$ with probability at least $1-\eta$. It uses $O(L\log{(nL/\eta)})$ maximum-flow computations, each on a network with $O(m+n)$ vertices and $O((m+n)\log^2{n})$ arcs with positive integer capacities bounded by $(n(W+1))^{O(1)}$, and  
    \[O((m+n)\log^2{n}\log{(m+n)}L\log{(nL/\eta)})\] additional expected arithmetic operations. If the  capacities of $G$ are polynomially bounded in $m+n$, then for $\eta=n^{-a}$ with any fixed $a>0$, the expected run-time is $(m+n)^{1+o(1)}$. 
\end{theorem}
We emphasize that our theorem is a reduction to max-flow. If max-flow can be done in near-linear time then the run-time of our algorithm also improves to near-linear. 

Several consequences follow directly. The root label gives the strength and its floor gives partition connectivity while the largest internal-node label gives fractional arboricity and its ceiling gives arboricity. These values can be extracted in $O(n)$ time. We therefore obtain almost linear-time algorithms for the integral and fractional tree packing numbers and for arboricity and fractional arboricity. 
Almost linear-time computation of arboricity was obtained only recently \cite{FLSW26}, while the previous fastest run-time for strength, partition connectivity, and the cut hierarchy was $nm^{1+o(1)}$ \cite{Bar92, CFLLP25}. 
As mentioned earlier, the hierarchy is a compact representation of the principal partition sequence and specifies the edge loads of a maximum fractional spanning tree packing. Thus, we get almost linear-time algorithms for an implicit representation of the principal partition sequence and a maximum fractional spanning tree packing. 

The main algorithmic ingredient of Theorem \ref{thm:cch} is an almost linear-time algorithm to solve the Lagrangian problem \eqref{eq:lagrangian} for graphs with polynomially bounded capacities as summarized in the following theorem. It is known that there exists a unique finest optimum partition to the Lagrangian problem \eqref{eq:lagrangian} \cite{Nar91, Nar97}. 

\begin{theorem}\label{thm:oracle}
Let $G=(V,E,c:E\to\mathbb Z_{+})$ be an undirected graph
with $n\ge2$ vertices and $m$ edges, and let
$\lambda=p/d\ge0$, where $p\ge0$ and $d\ge1$ are integers.
For every $\eta\in(0,1)$, there is a randomized algorithm
that returns the unique finest partition minimizing
$\Phi_{G,\lambda}$, together with its objective value,
with probability at least $1-\eta$. The algorithm has the following run-time guarantees, where $U:=\max\{c_e: e\in E\}$:
\begin{enumerate}
\item The algorithm uses $O(\log(n/\eta))$ maximum-flow
computations, each on a network with $O(m+n)$ vertices
and $O((m+n)\log^2 n)$ arcs, together with
\[
    O\!\left(
        (m+n)\log^2 n\log(m+n)\log(n/\eta)
    \right)
\]
additional expected arithmetic operations.
The maximum-flow networks can be chosen to have positive
integer capacities bounded by $\bigl(n(U+1)(p+d+1)\bigr)^{O(1)}$.

\item Moreover, let $S\ge m+n$ and let $C>0$ be fixed.
If $U,p,d, 1/\eta\le S^C$, then, provided 
each maximum-flow computation returns an integral maximum
flow, the additional expected running time is
\[
    O\!\left((m+n)\log^{O(1)}S\right),
\]
where the exponent depends only on $C$.

\item In particular, if the capacities and the integers $p,d$
are polynomially bounded in $m+n$, then, for
$\eta=n^{-a}$ with any fixed $a>0$, the expected running
time is $(m+n)^{1+o(1)}$.
\end{enumerate}
\end{theorem}

We remark that if max-flow can be done in near-linear time, then both of our algorithms also improve to near-linear time. 

\subsection{Technical Overview and AI Disclosure}\label{sec:techniques}
We first discuss our techniques to prove Theorem~\ref{thm:oracle}. 
Cut-respecting trees have been central to algorithms for global minimum cut. In Karger's algorithm, tree packing gives a small collection of trees such that a minimum cut crosses at most two edges of one of the trees in the collection; an efficient search over cuts with this property then recovers a minimum cut~\cite{Kar00}. This high-level principle generalizes for several other cut minimization problems and involves designing fast algorithms for the following two steps: (1) find a tree that crosses the relevant optimum cut a small number of times and (2) recover the relevant optimum/approximate-optimum cut from such a tree. This high-level principle has found applications in several other problems---e.g., fast computation of Gomory--Hu trees~\cite{AKLPST22, ALPS23, Zha21, AKLPGSYY25}, $k$-cut \cite{Tho08, CQX20, LSS24, CW23}, and directed global min-cut~\cite{FLSW26}. 
Our algorithm adopts the same high-level principle to solve the Lagrangian partition problem.

For ease of discussion in this technical section, we will assume that edge capacities and the numerator and denominator of $\lambda$ are polynomially bounded, and that there exists a unique optimum partition $\Pi^*$ for the Lagrangian problem \eqref{eq:lagrangian}. Throughout, a \emph{tree} will refer to a spanning tree over the vertex set $V$, and the tree need not be a subgraph of the input graph $G$. 
Every tree has at least $|\Pi^*|-1$ edges crossing $\Pi^*$. We say that a tree $T$ is \emph{$\Pi^*$-respecting} if it has exactly $|\Pi^*|-1$ such edges. 
Following the above-mentioned high-level principle, we show the following two results (stated informally): 

\begin{enumerate}[label={(\arabic*)}]
    \item There exists a randomized almost-linear time algorithm that returns a $\Pi^*$-respecting tree with high probability.
    \item Given a $\Pi^*$-respecting tree $T$, there exists a deterministic almost-linear time algorithm to find an optimum partition $\Pi$ for the Lagrangian problem. 
\end{enumerate}

The above two statements immediately lead to an almost-linear time algorithm to compute an optimum partition for the Lagrangian problem. We now outline our approach to prove the above two statements. We fix a root vertex $r\in V$ arbitrarily and retain this choice for the proofs of both statements. Both proofs rely on a network constructed relative to a fixed tree $T$. The proof of the second statement is easier to describe and motivates the network, so we begin with this. 

\paragraph{Recovering an optimum from a $\Pi^*$-respecting tree.}
We outline how to recover an optimum partition from a $\Pi^*$-respecting tree in almost-linear time. 
Fix $\lambda>0$ and normalize the input capacities by setting $q_e=c_e/\lambda$. It is convenient to maximize the \emph{partition deficit}
\begin{align*}
 D(\Pi)&:= |\Pi|-1-q(\delta_G(\Pi)),\\
 D^*&:= \max_{\Pi}D(\Pi)=-\opt_G(\lambda)/\lambda.
\end{align*}
This is exactly the Lagrangian partition problem \eqref{eq:lagrangian} after multiplying the objective by $-1/\lambda$. The trivial partition containing only one part has deficit zero, so $D^*\ge0$. 

We construct a network $\mathcal{N}(T)$ that depends on the tree $T$. The network $\mathcal N(T)$ has one unit of supply associated with each nonroot vertex and one unit of receiving capacity associated with each tree edge. There are $n-1$ of each. To send a unit from a vertex $v$ to a tree edge $f$, the flow must choose a graph edge $e=vu$ incident to $v$ such that the unique $v$--$u$ path in $T$ contains $f$. Thus, a flow path records both a selected graph edge and a tree edge lying on its tree path. All flow paths using the same graph edge $e$ pass through a shared arc of capacity $q_e$, including paths starting from either endpoint. This prevents the two end-vertices of $e$ from using the capacity of $e$ independently. 
See Section~\ref{sec:flow} for the full network and its compact implementation, which has near-linear size and permits an almost-linear-time maximum-flow computation~\cite{BCKLPPSS23}.

The maximum flow in the network can be at most $n-1$ by construction. 
We write $f_T$ for the maximum-flow value in this network and $\alpha_T=n-1-f_T$ for its \emph{shortfall}. The network has two complementary cut properties. First, from any minimum cut we can extract a partition whose deficit is at least $\alpha_T$. The extraction uses the source-side vertex nodes and tree-edge nodes: certain components of the corresponding tree-edge forest become parts, and the remaining vertices form the part containing $r$. The shared input-edge arcs pay for the edges crossing this partition. Consequently $\alpha_T\le D^*$ for every tree $T$ irrespective of whether the tree respects an optimum partition.

Second, if $T$ respects a partition $\Pi$, its parts are subtrees and can be encoded by a network cut of capacity $n-1-D(\Pi)$. The part containing the arbitrarily chosen root plays a distinguished role in this encoding. If $\Pi$ is optimal, this gives $\alpha_T\ge D^*$. The two inequalities then force $\alpha_T=D^*$, and the partition extracted from a minimum cut is optimal. Crucially, constructing the network and extracting this partition use only $G$, $T$, and $\lambda$. The unknown respecting partition is needed solely to prove that the recovered partition is optimal. We formalize these details in Section \ref{sec:opt-from-respecting-tree}. 

\paragraph{Finding a $\Pi^*$-respecting tree.} 
Next, we outline how to find a $\Pi^*$-respecting tree without knowledge of $\Pi^*$. 
Define the excess of a tree $T$ by
\[
 z_{\Pi^*}(T)=|T\cap\delta(\Pi^*)|-(|\Pi^*|-1),
\]
The excess is a nonnegative integer, and it is zero exactly when $T$ respects $\Pi^*$. Given an arbitrary tree $T$, we will design a randomized algorithm to construct an updated tree $T'$ satisfying
\begin{align}
 \E[z_{\Pi^*}(T')]&\le\tfrac12 z_{\Pi^*}(T). \label{ineq:halving}
\end{align}
Starting from any tree on $V$, after $O(\log(n/\eta))$ such updates, the expected excess is at most $\eta$. Its integrality implies that the final tree respects $\Pi^*$ with probability at least $1-\eta$. 

We now explain how to construct a tree $T'$ satisfying \eqref{ineq:halving}. We will use the max flow in the network $\mathcal{N}(T)$. 
For ease of exposition, we focus on the case where we have a flow of value $n-1$ in the network here (if its value is less than $n-1$, we first complete the flow through an auxiliary node; this may introduce selected edges outside $G$, which is a mild technical complication---Section~\ref{sec:halving-algo} for details). Randomized flow rounding~\cite{KP15} gives an integral flow while preserving the expected flow on every arc. Such an integral flow corresponds to unit paths. These unit paths would select one graph edge $e=uv$ for each nonroot vertex $v$ and assign the selected edges bijectively to the tree edges, with each assigned tree edge $f$ lying on the unique tree-path between the end-vertices of the selected graph edge $e=uv$ being assigned to $f$. The shared capacity arc in the network would ensure that the expected number of selected copies of a graph edge $e$ is at most $q_e$. 
Let $F$ be the resulting multiset of $n-1$ selected graph edges. The bijective assignment and path condition imply that the multigraph $E(T)\uplus F$ decomposes into two spanning trees and such a decomposition can be found in $O(n\log n)$ time~\cite{Tar76}; see Lemma~\ref{lem:two-tree-decomposition}. Choose one of these two trees uniformly to obtain $T'$.

We now outline why \eqref{ineq:halving} holds for such a randomized choice of $T'$. First, we bound how many selected copies in $F$ cross the fixed optimal partition $\Pi^*$. Let $X_e$ be the number of copies of graph edge $e$ selected into $F$. Each such copy uses the shared capacity arc for $e$. Unbiased rounding preserves its expected flow, so $\E[X_e]\le q_e$. Only copies of graph edges in $\delta_G(\Pi^*)$ contribute original crossings. Linearity of expectation therefore gives
\[
 \E[|F\cap \delta(\Pi^*)|]
 =\sum_{e\in\delta_G(\Pi^*)}\E[X_e]
 \le q(\delta_G(\Pi^*))=(|\Pi^*|-1)-D^* \le |\Pi^*|-1.
\]
For each realized $F$, the two new trees together contain exactly the crossing copies in $E(T)\uplus F$. 
The current tree $T$ has $(|\Pi^*|-1)+z_{\Pi^*}(T)$ crossings, whereas $F$ contributes at most $|\Pi^*|-1$ in expectation. Consequently, uniform choice of a new tree $T'$ gives
\[
 \E[|T'\cap\delta(\Pi^*)|]
 =\tfrac12\bigl(|T\cap\delta(\Pi^*)|
 +\E[|F\cap\delta(\Pi^*)|]\bigr)
 \le (|\Pi^*|-1)+\tfrac12 z_{\Pi^*}(T).
\]
Subtracting $|\Pi^*|-1$ proves that the expected excess halves. We emphasize that the argument controls the crossings above the unavoidable $|\Pi^*|-1$, rather than seeking to halve the number of crossings. We formalize these details in Section \ref{sec:find-respecting-tree}. 

\paragraph{Comparison to known techniques.}
Our work and techniques build on the recent work on showing an almost linear-time algorithm for directed global min-cut \cite{FLSW26}. Their algorithm is also based on the high-level principle of finding a cut-respecting tree and working with such a tree to find an optimum cut. They use a network for both steps and that network informs the construction of our network, although our network is different from theirs. 
A distinction of our work from all previous appearances of the high-level principle 
is the number of relevant crossing tree edges. All previous algorithms were interested in cuts, which is a partition of the vertex set into two parts, and used trees with only a small number of edges crossing the optimum cut of interest. For the Lagrangian problem that is of interest to our work, all optimum partitions could have a large number of parts. Consequently, we cannot afford to use trees that have only a small number of edges crossing an optimum partition $\Pi^*$. 
In our setting, a $\Pi^*$-respecting tree has exactly $|\Pi^*|-1$ edges crossing the optimum partition, and this number can grow with $n$. Thus, our work shows the generalizability of the approach beyond a constant number of crossings.

\paragraph{Canonical cut hierarchy from the Lagrangian.}
We now explain how Theorem \ref{thm:oracle} implies Theorem \ref{thm:cch}. The principal partition sequence provides the link: at each $\lambda$, the optimum partition divides the hierarchy into splits that have already occurred and splits that remain to be discovered. We recover these two portions by contraction and restriction. For simplicity assume that $G$ is connected. Let $W:=\sum_{e\in E}c_e$. Start with interval $[0, W+1]$. The finest optima to the Lagrangian problem at the end-points of this interval are the one-part and singleton partitions, respectively. A recursive call on a graph $H$ and an interval $[\ell, u]$ maintains these end-point optima. Find the finest optimum partition $\Pi$ to the Lagrangian problem at the mid-point $t:=(\ell+u)/2$. Next, observe that the finest optimum partitions for the Lagrangian problem for $\lambda$ values below $t$ can only merge parts of $\Pi$. To find these, we contract each part of $\Pi$ to a vertex and recurse on the resulting graph $H/\Pi$ over the interval $[\ell, t]$. Next, observe that the finest optimum partitions for the Lagrangian problem for $\lambda$ values above $t$ can only split parts of $\Pi$. To find the later splits, recurse separately on each induced graph $H[B]$ for each $B\in \Pi$ over the interval $[t, u]$. The recursion on $H/\Pi$ returns a hierarchy whose leaves represent the parts of $\Pi$. Replace the leaf representing $B$ by the hierarchy returned for $H[B]$. This way, we assemble the full hierarchy without explicitly listing every partition in the sequence. We consider two base cases for the recursion: (1) If $H$ is a one-vertex graph, we return a single leaf. (2) If the interval width $|u-\ell|$ is less than $1/(n-1)^2$; integer capacities ensure that the interval contains at most one breakpoint. Since the end-point partitions for this interval differ, it contains exactly one breakpoint at which the one-part and singleton objectives agree. The resulting hierarchy is a split of $V(H)$ into singletons with label $c_H(E(H))/(|V(H)|-1)$. See Section \ref{sec:cch} for details. 

The running time follows from conservation of total graph size at each recursion depth. Every edge of $H$ appears in exactly one child problem: edges crossing $\Pi$ go to the recursion problem on $H/\Pi$ and edges within a part go to that part's induced graph. Consequently, the instances at each depth of the recursion tree have $O(m)$ total size. Consequently, the total expected run-time at each depth is $m^{1+o(1)}$. We will reach the base case of the recursion in at most $O(\log{(n(W+1))})$ levels. Hence, the total expected run-time is almost-linear. The use of nesting and restriction-contraction principles to construct the principal partition sequence is classical\cite{Nar91}. We use the same principles to obtain an almost linear time algorithm via the combination of almost linear-time finest Lagrangian optimum partition computation, conservation of total instance size at each level, and a logarithmic bound on the number of levels.

\paragraph{AI Disclosure.} This paper originated as follows: the authors had a faster algorithm for arboricity in simple unweighted graphs via new structural observations. While writing up those results, the work on directed global min-cut \cite{FLSW26} was posted to arXiv. One of the consequences of that work is an almost linear-time algorithm to compute arboricity. The authors asked AI to use the ideas in that paper to design an almost linear-time algorithm for computing strength, which it promptly did. The authors noticed that the ideas to compute strength would generalize to compute the canonical cut hierarchy in almost linear time. The authors asked AI to complete this step as well and verified the details. We used AI to assist with writing this manuscript. We assume responsibility for all content.

\subsection{Preliminaries}
Let $G=(V, E, q: E\rightarrow \mathbb{Q}_+)$ be a graph with positive rational capacities. Throughout this work, a tree refers to a spanning tree over vertex set $V$, a partition refers to a partition of the vertex set $V$ into non-empty parts, and the Lagrangian problem on $G$ refers to the Lagrangian problem \eqref{eq:lagrangian} on $G$ with parameter $\lambda=1$. 
Let $T$ be a tree over $V$. 
For vertices $u, v\in V$, let $P_T(u, v)$ be the edge set of the unique $u-v$ path in $T$. 
Let $\Pi$ be a partition. 
We denote $\delta_G(\Pi)$ to be the set of edges in $G$ whose end-vertices are in distinct parts of $\Pi$ and $\delta_T(\Pi)$ to be the set of edges in $T$ whose end-vertices are in distinct parts of $\Pi$. 
We say that $T$ is $\Pi$-respecting if the tree has exactly $|\Pi|-1$ edges crossing $\Pi$. 
We define the \emph{deficit} of $\Pi$ and the maximum deficit to be 
\begin{align*}
 D(\Pi)&:=|\Pi|-1-q(\delta_G(\Pi)) \text{ and}\\
 D^*&:=\max_{\Pi} D(\Pi).
\end{align*}
We note that the Lagrangian problem \eqref{eq:lagrangian} on $G$ is equivalent to finding a partition that maximizes deficit and hence, $\Pi^*$ is a deficit-maximizing partition. 
We denote $[b]:=\{1, 2, 3, \ldots, b\}$. 
\section{A tree-based network and its properties}\label{sec:flow}
Let $G=(V,E,q:E\to\mathbb{Q}_{>0})$ be an undirected graph with
positive rational capacities, where $n:=|V|$ and $m:=|E|$.
Fix an arbitrary root vertex $r\in V$.
We construct a network that will be used both to recover an
optimum Lagrangian partition from a respecting tree and to find
such a tree.

Let $T$ be a tree on vertex set $V$. The tree need not be a
subgraph of $G$, and its edges do not carry capacities.
For $u,v\in V$, let $P_T(u,v)$ denote the edge set of the unique
$u$--$v$ path in $T$. We write $\delta_G(v)$ for the set of graph
edges incident with $v$.

\begin{definition}\label{def:pps-network}
The directed capacitated network $\mathcal N(T)$ has source $s$,
sink $t$, and the following additional nodes:
\begin{enumerate}[label=\arabic*.,leftmargin=2em,itemsep=2pt,topsep=4pt]
\item a node $a_v$ for every $v\in V\setminus\{r\}$;
\item two nodes $y_e^-$ and $y_e^+$ for every graph edge $e\in E$;
\item a node $z_f$ for every tree edge $f\in E(T)$.
\end{enumerate}
Its arcs and their capacities are as follows:
\begin{enumerate}[label=\arabic*.,leftmargin=2em,itemsep=2pt,topsep=4pt]
\item an arc $s\to a_v$ of capacity $1$, for every
      $v\in V\setminus\{r\}$;
\item an arc $z_f\to t$ of capacity $1$, for every $f\in E(T)$;
\item an arc $y_e^-\to y_e^+$ of capacity $q_e$, for every $e\in E$;
\item an arc $a_v\to y_e^-$ of capacity $\infty$, for every
      $v\in V\setminus\{r\}$ and every $e\in\delta_G(v)$;
\item an arc $y_e^+\to z_f$ of capacity $\infty$, for every
      $e=uv\in E$ and every $f\in P_T(u,v)$.
\end{enumerate}
We call the arcs in the last item the \emph{path-incidence arcs}.
Let $f_T$ be the maximum $(s,t)$-flow value in $\mathcal N(T)$,
and define its \emph{shortfall} by
\[
    \alpha_T:=n-1-f_T.
\]
We identify an $(s,t)$-cut with its source side $A$, where
$s\in A$ and $t\notin A$. Its capacity, denoted
$\operatorname{cap}_{\mathcal N(T)}(A)$, is the sum of the
capacities of the arcs leaving $A$.
\end{definition}

\begin{figure}[htbp]
\centering
\begin{tikzpicture}[
  >=Stealth,
  networknode/.style={draw,rounded corners=2pt,minimum width=7mm,
                       minimum height=7mm,inner sep=2pt},
  terminal/.style={draw,circle,inner sep=2pt},
  every node/.style={font=\small},
  every edge/.style={draw,->,line width=.65pt}
]
\node[terminal] (s) at (0,0) {$s$};
\node[networknode] (au) at (1.8,.85) {$a_u$};
\node[networknode] (av) at (1.8,-.85) {$a_v$};
\node[networknode] (ym) at (3.6,0) {$y_e^-$};
\node[networknode] (yp) at (5.4,0) {$y_e^+$};
\node[networknode] (z1) at (7.2,.85) {$z_{f_1}$};
\node[networknode] (z2) at (7.2,-.85) {$z_{f_2}$};
\node[terminal] (t) at (9,0) {$t$};
\path (s) edge node[above left] {$1$} (au);
\path (s) edge node[below left] {$1$} (av);
\path (au) edge node[above right] {$\infty$} (ym);
\path (av) edge node[below right] {$\infty$} (ym);
\path (ym) edge node[above] {$q_e$} (yp);
\path (yp) edge node[above left] {$\infty$} (z1);
\path (yp) edge node[below left] {$\infty$} (z2);
\path (z1) edge node[above right] {$1$} (t);
\path (z2) edge node[below right] {$1$} (t);
\end{tikzpicture}
\caption{A portion of $\mathcal N(T)$ for a graph edge $e=uv$,
with $u,v\ne r$ and $f_1,f_2\in P_T(u,v)$. }
\label{fig:network}
\end{figure}
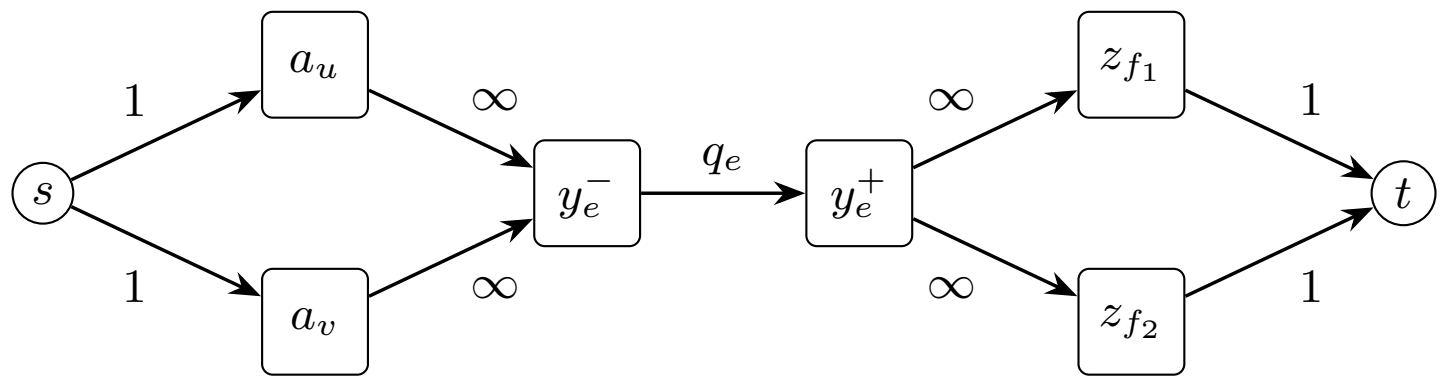

Figure~\ref{fig:network} illustrates the construction.
The shared arc $y_e^-\to y_e^+$ limits the total flow routed
through a graph edge $e=uv$; after using this arc, flow can
proceed to any node $z_f$ with $f\in P_T(u,v)$.
The source cut $\{s\}$ has capacity $n-1$, so
\[
    0\le f_T\le n-1.
\]
In particular, $\alpha_T\ge0$.
The network is acyclic, and every feasible flow decomposes into
$(s,t)$-paths of total weight at most $n-1$.
Thus every arc carries at most $n-1$ units, and replacing all
infinite capacities by $n$ preserves every feasible flow.
This replacement also preserves, in both directions, every
$(s,t)$-cut of capacity at most $n-1$: no such cut can cross an
arc of capacity $\infty$ or $n$.
All algorithmic constructions below use this finite implementation.

\subsection{Compressing the network}\label{subsec:compact}
The network $\mathcal N(T)$ has $O(m+n)$ nodes, but its
path-incidence arcs can number $\Theta(mn)$.
We replace these arcs by a small directed network that preserves
which tree-edge nodes are reachable from each $y_e^+$.
The size bounds below are stated for $n\ge2$; the case $n=1$
has no tree edges and is immediate.

\begin{lemma}\label{lem:pps-compact-network}
Given an undirected graph $G=(V,E,q:E\to\mathbb{Q}_{>0})$
with $n\ge2$ vertices and $m$ edges, a tree $T$ on $V$, and
a root vertex $r\in V$, one can construct a directed acyclic
network $\compact(T)$ with the following properties.
It contains every node of $\mathcal N(T)$ and retains every
arc other than the path-incidence arcs, with infinite
capacities implemented by $n$.
\begin{enumerate}[label=\arabic*.,leftmargin=2em,itemsep=5pt,topsep=5pt]
\item The network has $O(m+n)$ nodes and $O(n+m\log^2 n)$ arcs
      and can be constructed in $O(n+m\log^2 n)$ time.
      Every arc capacity belongs to
      \[
          \{1,n\}\cup\{q_e:e\in E\}.
      \]
\item The maximum $(s,t)$-flow value in $\compact(T)$ is $f_T$.
\item Every $(s,t)$-cut with source side $A_{\mathrm{compact}}$
      in $\compact(T)$ and capacity at most $n-1$ restricts to
      an $(s,t)$-cut with source side
      \[
          A=A_{\mathrm{compact}}\cap V(\mathcal N(T))
      \]
      of the same capacity in $\mathcal N(T)$.
      Conversely, every $(s,t)$-cut in $\mathcal N(T)$ of
      capacity at most $n-1$ extends to an $(s,t)$-cut of the
      same capacity in $\compact(T)$.
\item For every graph edge $e=uv\in E$ and every tree edge
      $f\in E(T)$,
      \[
          y_e^+\text{ reaches }z_f\text{ in }\compact(T)
          \quad\Longleftrightarrow\quad
          f\in P_T(u,v).
      \]
      Each such directed path consists of one arc from
      $y_e^+$ to a node of an auxiliary rooted forest,
      followed by zero or more arcs from forest parents to
      their children. Its length is $O(\log n)$.
\end{enumerate}
\end{lemma}

The construction uses the following representation of tree paths.
It combines a heavy-light decomposition~\cite{ST83} with balanced
binary trees on the heavy paths. We give the full construction
and query algorithm.

\begin{lemma}\label{lem:pps-tree-path-forest}
Given a tree $T=(V,E(T))$ with $n\ge2$ vertices, one can
construct in $O(n)$ time a rooted binary forest $\mathcal B$
and auxiliary data of total size $O(n)$ with the following
properties.
\begin{enumerate}[label=\arabic*.,leftmargin=2em,itemsep=5pt,topsep=5pt]
\item The forest has $O(n)$ nodes and height $O(\log n)$.
      Its leaves are in bijection with $E(T)$: for each
      $f\in E(T)$ there is exactly one leaf $\ell_f$, and
      every leaf is of this form.
      For each node $x$ of $\mathcal B$, define
      \[
          L(x):=\{f\in E(T):\ell_f\text{ lies in the subtree
          of }\mathcal B\text{ rooted at }x\}.
      \]
      In particular, $L(\ell_f)=\{f\}$.
\item For every $u,v\in V$, one can compute in $O(\log^2 n)$
      time a set $C(u,v)$ of $O(\log^2 n)$ nodes of
      $\mathcal B$ such that the sets $L(x)$, for
      $x\in C(u,v)$, are pairwise disjoint and
      \[
          P_T(u,v)=\biguplus_{x\in C(u,v)}L(x).
      \]
      When $u=v$, the returned set is empty.
\end{enumerate}
\end{lemma}

\begin{proof}
Root $T$ at an arbitrary vertex, and compute the size $s(v)$
of the subtree rooted at each vertex $v$.
For every vertex with children, choose a child of maximum
subtree size, breaking ties arbitrarily, and call its incident
edge \emph{heavy}; call every other parent--child edge
\emph{light}.
If $vw$ is light and $v$ is the parent of $w$, a different
child of $v$ has subtree size at least $s(w)$, so
\[
    2s(w)\le s(v)-1.
\]
Consequently, a path from the root to any vertex contains
$O(\log n)$ light edges.
The heavy edges form vertex-disjoint paths, including
single-vertex paths. The \emph{head} of such a path is its
vertex nearest the root.
Store the parent, depth, path head, and position on its heavy
path for every vertex. All these data can be computed in
$O(n)$ time.

For each heavy path with at least one edge, order its edges
from its head downwards and build a balanced binary tree
whose leaves, in that order, correspond to these edges.
Construct this binary tree by recursively dividing the edge
list into two consecutive sublists whose lengths differ
by at most one, and store the first and last leaf positions
at every binary-tree node.
For each light edge $f$, create a separate one-node tree
consisting of the leaf $\ell_f$.
Let $\mathcal B$ be the disjoint union of these binary trees.
Every edge of $T$ labels exactly one leaf.
A binary tree on $k$ leaves has $2k-1$ nodes and height
$O(\log(k+1))$ in this construction.
Since the leaf sets partition $E(T)$, the forest has $O(n)$
nodes, height $O(\log n)$, and can be built in $O(n)$ time.
We store its nodes and leaf labels; the sets $L(x)$ are
represented implicitly by the subtrees.

We next describe how to answer a query $(u,v)$, i.e. given two vertices $u,v\in V$, we show
to compute the set $C(u,v)$ required by Property~2. Maintain two current vertices, initially $u$ and $v$.
While their heavy-path heads differ, select the current
vertex whose head is deeper, breaking ties arbitrarily.
Report the heavy-edge interval from that head to the
selected vertex, if nonempty, and the light edge entering
the head. Replace the selected vertex by the parent of
its head. When the two current vertices lie on the same
heavy path, report the interval between them, if nonempty,
and stop.
The chosen head is below the lowest common ancestor of the
two current vertices, so every reported interval or edge
lies on their connecting path. Each step removes precisely
that portion of the path. Thus the reported intervals
and light edges are pairwise edge-disjoint and together
equal $P_T(u,v)$.
Each iteration crosses one light edge on one of the two
original paths towards their lowest common ancestor.
Hence there are $O(\log n)$ iterations, and the stored
data allow each interval and light edge to be identified
in constant time per iteration.

It remains to represent a heavy-edge interval using nodes
of its binary tree in $\mathcal B$.
Starting at that tree's root, discard a node if its leaf
interval is disjoint from the queried interval, select it
if its leaf interval is contained in the queried interval,
and otherwise recurse on its children.
The selected subtrees have disjoint leaf sets whose union
is exactly the queried interval.
At each depth, at most two nodes have leaf intervals that
intersect the query without being contained in it.
The search therefore visits and selects $O(\log n)$ nodes
and takes $O(\log n)$ time.
Apply this procedure to every reported heavy-edge interval,
and select the singleton leaf for every reported light edge.
The union of all selected nodes is $C(u,v)$.
There are $O(\log n)$ heavy-edge intervals and light edges, so the
query takes $O(\log^2 n)$ time and returns $O(\log^2 n)$
nodes. The disjointness and covering properties follow
from those of the reported path portions and the selected
binary-tree subtrees.
\end{proof}

We now prove Lemma~\ref{lem:pps-compact-network}.

\begin{proof}[Proof of Lemma~\ref{lem:pps-compact-network}]
We first describe the construction, using the finite
implementation of $\mathcal N(T)$ described above.
Construct $\mathcal B$ using
Lemma~\ref{lem:pps-tree-path-forest}.
Identify each leaf $\ell_f$ with the existing node $z_f$.
Add the internal nodes of $\mathcal B$ to the network,
and direct every forest edge from parent to child.
For each graph edge $e=uv\in E$, compute
$C_e:=C(u,v)$ and add the arcs
\[
    y_e^+\longrightarrow x
    \qquad\text{for every }x\in C_e.
\]
Give every forest arc and every added arc capacity $n$.
Retain all other nodes and arcs of $\mathcal N(T)$,
omitting the original path-incidence arcs.
Call the resulting network $\compact(T)$.

We next verify the preservation properties.
The network is acyclic: its arcs proceed from $s$ to
the $a_v$ nodes, then to the $y_e^-$ nodes, then to the
$y_e^+$ nodes, and finally down the forest to the $z_f$
nodes and $t$.
For every graph edge $e=uv$, the forest leaves reachable
from $y_e^+$ are exactly the leaves below the nodes in
$C_e$. By Lemma~\ref{lem:pps-tree-path-forest}, their
labels are precisely the edges of $P_T(u,v)$.
This proves the reachability equivalence in part~4.
Every such directed path uses one added arc followed by
a downward forest path, so its length is $O(\log n)$.

Let $A_{\mathrm{compact}}$ be the source side of an
$(s,t)$-cut in $\compact(T)$ with capacity at most $n-1$,
and set $A:=A_{\mathrm{compact}}\cap V(\mathcal N(T))$.
No added arc or forest arc leaves $A_{\mathrm{compact}}$,
since each has capacity $n$.
Thus, if $y_e^+\in A$, every forest leaf reachable from
$y_e^+$ also belongs to $A$.
The reachability equivalence implies that no omitted
path-incidence arc leaves $A$.
All remaining arcs are retained and cross the two cuts
identically, with the same finite capacities.
Therefore $A$ is the source side of an $(s,t)$-cut in
$\mathcal N(T)$ of the same capacity.

Conversely, let $A$ be the source side of an $(s,t)$-cut
in $\mathcal N(T)$ with capacity at most $n-1$.
Extend $A$ by adding every internal forest node reachable
from some $y_e^+\in A$, and call the resulting set
$A_{\mathrm{compact}}$.
Every forest leaf reached in this way already belongs
to $A$: otherwise a corresponding path-incidence arc
would leave $A$ and contribute capacity $n$.
It follows that no added arc or forest arc leaves
$A_{\mathrm{compact}}$.
No omitted path-incidence arc leaves $A$, and all
retained arcs cross the two cuts identically.
This proves the reverse cut correspondence and hence
part~3.
Both networks have a source cut of capacity $n-1$.
Applying part~3 to a minimum cut in each network shows
that their minimum-cut values agree.
The max-flow/min-cut theorem now gives part~2.

Finally, we bound the construction time and network size.
The retained portion of $\mathcal N(T)$ has $O(m+n)$
nodes and arcs. The forest adds $O(n)$ nodes and arcs.
For each graph edge, its query and the addition of its
$O(\log^2 n)$ arcs take $O(\log^2 n)$ time.
Hence the compact network has $O(m+n)$ nodes and
$O(n+m\log^2 n)$ arcs, and it is constructed in
$O(n+m\log^2 n)$ time.
Every retained arc has capacity $1$, $q_e$ for some
$e\in E$, or $n$, and every new arc has capacity $n$.
This proves all the claims in part~1.
\end{proof}

\section{Optimal partition from a respecting tree}\label{sec:opt-from-respecting-tree}
Let $G=(V,E,q:E\to\mathbb{Q}_{>0})$ be an undirected graph with positive rational capacities and a fixed root vertex $r\in V$, where $n:=|V|$ and $m:=|E|$. 
Let $\Pi^*$ be an optimum partition for the Lagrangian problem. In this section, we show that a $\Pi^*$-respecting tree can be used to find an optimum partition for the Lagrangian problem using a single max-flow computation in a network of near-linear size. The following lemma is the main result of this section. 

\begin{lemma}
\label{lem:pps-recover-optimum}
Given an undirected graph
$G=(V,E,q:E\rightarrow\mathbb{Q}_+)$ with $n\ge2$ vertices
and $m$ edges, and a $\Pi^*$-respecting tree $T$ for some
optimum partition $\Pi^*$ to the Lagrangian problem on $G$,
there exists a deterministic algorithm that returns an
optimal partition to the Lagrangian problem on $G$.
The algorithm has the following running-time guarantees:
\begin{enumerate}
\item The algorithm uses one maximum-flow computation on
a network with $O(m+n)$ vertices and
$O((m+n)\log^2 n)$ arcs, where every arc capacity belongs to
$\{1,n\}\cup\{q_e:e\in E\}$, together with
$O((m+n)\log^2 n)$ additional arithmetic operations.

\item Let $S\ge m+n$ and let $C>0$ be fixed.
Suppose that $q_e=a_e/B$ for every $e\in E$, where $B$
and all $a_e$ are positive integers satisfying
$B\le S^C$ and $a_e\le S^C$ for every $e\in E$.
If the maximum-flow computation returns a flow whose arc
values are integer multiples of $1/B$, then the additional
running time is
\[
    O\!\left((m+n)\log^{O(1)}S\right),
\]
where the exponent depends only on $C$.
\end{enumerate}

\end{lemma}

We first show two intermediate lemmas that help us establish Lemma~\ref{lem:pps-recover-optimum}. We recall that our goal is to find $D^*$ and a deficit-maximizing partition. 
Our first lemma shows that if $T$ is a $\Pi^*$-respecting tree, then shortfall is at least the maximum deficit.

\begin{lemma}
\label{lem:pps-respecting-cut}
Let $T$ be a $\Pi^*$-respecting tree for some optimum partition $\Pi^*$ to the Lagrangian problem, then
$\mathcal N(T)$ has an $(s,t)$-cut of capacity $n-1-D^*$.
Consequently, $\alpha_T\ge D^*$.
\end{lemma}

\begin{proof}
Let $\Pi^*=\{P_0,\ldots,P_{k-1}\}$ with vertex $r\in P_0$.
Since exactly $k-1$ tree edges cross $\Pi^*$, the induced
forests $T[P_i]$ together contain $n-k$ edges.
Each contains at most $|P_i|-1$ edges, and
\[
 \sum_{i=0}^{k-1}(|P_i|-1)=n-k.
\]
Thus, every $T[P_i]$ is connected.
Define
\begin{align*}
 U^* &= V\setminus P_0,\\
 J^* &= \bigcup_{i=1}^{k-1}E(T[P_i]).
\end{align*}
Then, $|J^*|=|U^*|-(k-1)$.
Construct the set $A^*\subseteq V(\mathcal{N}(T))$ that contains the source vertex $s$ and the following vertices: 
\begin{enumerate}
 \item all $a_v$ with $v\in U^*$, and all $z_f$
       with $f\in J^*$, 
 \item $y_e^-$ for every edge $e\in E$ with an end-vertex in $U^*$, and 
 \item $y_e^+$ for every edge $e=uv\in E$ with $u, v\in P_i$ for some $i\in [k-1]$. 
\end{enumerate}

\begin{figure}[t]
\centering
\begin{tikzpicture}[
 x=1cm,y=1cm,font=\small,
 dot/.style={
     circle,draw=black,minimum size=8mm,inner sep=1pt
 },
 insidecut/.style={dot,fill=black!20,text=black},
 outsidecut/.style={dot,fill=white,text=black},
 part/.style={
     draw,densely dotted,rounded corners=5pt,inner sep=5pt
 },
 arc/.style={-{Latex[length=2mm]},line width=0.45pt},
 cutarc/.style={arc,line width=1.3pt},
 captext/.style={
     fill=white,text=black,inner sep=1pt,font=\scriptsize
 }
]

\node[outsidecut] (tr) at (0.8,5.4) {$r$};
\node[outsidecut] (t1) at (2.5,5.4) {$v_1$};
\node[insidecut]  (t2) at (4.2,5.4) {$v_2$};
\node[insidecut]  (t3) at (5.9,5.4) {$v_3$};
\node[insidecut]  (t4) at (7.6,5.4) {$v_4$};
\node[insidecut]  (t5) at (9.6,5.4) {$v_5$};

\draw (tr)--node[above] {$e_1$}
            node[below] {$2$}(t1);
\draw[dashed] (t1)--node[above] {$e_2$}
                    node[below] {$\frac12$}(t2);
\draw[line width=1.3pt] (t2)--node[above] {$e_3$}
                             node[below] {$2$}(t3);
\draw[line width=1.3pt] (t3)--node[above] {$e_4$}
                             node[below] {$2$}(t4);
\draw[dashed] (t4)--node[above] {$e_5$}
                    node[below] {$\frac12$}(t5);

\node[part,fit=(tr)(t1),label=above:{$P_0$}] {};
\node[part,fit=(t2)(t3)(t4),label=above:{$P_1$}] {};
\node[part,fit=(t5),label=above:{$P_2$}] {};

\node at (5.2,4.25) {(a)};

\draw[dashed,rounded corners=4pt,line width=0.7pt]
    (-0.65,1.8)--(5.5,1.8)--(5.5,0.6)--(9.05,0.6)
    --(9.05,-1.8)--(5.5,-1.8)--(5.5,-3)
    --(-0.65,-3)--cycle;
\node at (0.15,1.4) {$A^*$};

\node[insidecut]  (s) at (0,0) {$s$};
\node[outsidecut] (t) at (10.8,0) {$t$};

\node[outsidecut] (a1) at (2,2.4)  {$a_{v_1}$};
\node[insidecut]  (a2) at (2,1.2)  {$a_{v_2}$};
\node[insidecut]  (a3) at (2,0)    {$a_{v_3}$};
\node[insidecut]  (a4) at (2,-1.2) {$a_{v_4}$};
\node[insidecut]  (a5) at (2,-2.4) {$a_{v_5}$};

\node[outsidecut] (ym1) at (4.3,2.4)  {$y_{e_1}^-$};
\node[insidecut]  (ym2) at (4.3,1.2)  {$y_{e_2}^-$};
\node[insidecut]  (ym3) at (4.3,0)    {$y_{e_3}^-$};
\node[insidecut]  (ym4) at (4.3,-1.2) {$y_{e_4}^-$};
\node[insidecut]  (ym5) at (4.3,-2.4) {$y_{e_5}^-$};

\node[outsidecut] (yp1) at (6.5,2.4)  {$y_{e_1}^+$};
\node[outsidecut] (yp2) at (6.5,1.2)  {$y_{e_2}^+$};
\node[insidecut]  (yp3) at (6.5,0)    {$y_{e_3}^+$};
\node[insidecut]  (yp4) at (6.5,-1.2) {$y_{e_4}^+$};
\node[outsidecut] (yp5) at (6.5,-2.4) {$y_{e_5}^+$};

\node[outsidecut] (z1) at (8.7,2.4)  {$z_{e_1}$};
\node[outsidecut] (z2) at (8.7,1.2)  {$z_{e_2}$};
\node[insidecut]  (z3) at (8.7,0)    {$z_{e_3}$};
\node[insidecut]  (z4) at (8.7,-1.2) {$z_{e_4}$};
\node[outsidecut] (z5) at (8.7,-2.4) {$z_{e_5}$};

\draw[cutarc] (s)--node[captext,above] {$1$}(a1);
\draw[arc]    (s)--node[captext,above] {$1$}(a2);
\draw[arc]    (s)--node[captext,above] {$1$}(a3);
\draw[arc]    (s)--node[captext,above] {$1$}(a4);
\draw[arc]    (s)--node[captext,above] {$1$}(a5);

\draw[arc] (a1)--(ym1);
\draw[arc] (a1)--(ym2);
\draw[arc] (a2)--(ym2);
\draw[arc] (a2)--(ym3);
\draw[arc] (a3)--(ym3);
\draw[arc] (a3)--(ym4);
\draw[arc] (a4)--(ym4);
\draw[arc] (a4)--(ym5);
\draw[arc] (a5)--(ym5);

\draw[arc]    (ym1)--node[captext,above] {$2$}(yp1);
\draw[cutarc] (ym2)--node[captext,above] {$\frac12$}(yp2);
\draw[arc]    (ym3)--node[captext,above] {$2$}(yp3);
\draw[arc]    (ym4)--node[captext,above] {$2$}(yp4);
\draw[cutarc] (ym5)--node[captext,above] {$\frac12$}(yp5);

\draw[arc] (yp1)--(z1);
\draw[arc] (yp2)--(z2);
\draw[arc] (yp3)--(z3);
\draw[arc] (yp4)--(z4);
\draw[arc] (yp5)--(z5);

\draw[arc]    (z1)--node[captext,above] {$1$}(t);
\draw[arc]    (z2)--node[captext,above] {$1$}(t);
\draw[cutarc] (z3)--node[captext,above] {$1$}(t);
\draw[cutarc] (z4)--node[captext,above] {$1$}(t);
\draw[arc]    (z5)--node[captext,above] {$1$}(t);

\node at (5.2,-3.55) {(b)};

\end{tikzpicture}
\caption{
An example of the construction in
Lemma~\ref{lem:pps-respecting-cut}.
(a) The input graph $G$ is the path
$r-v_1-v_2-v_3-v_4-v_5$, and the chosen tree $T$
is this same path.
Dotted outlines show the parts of
$\Pi^*=\{\{r,v_1\},\{v_2,v_3,v_4\},\{v_5\}\}$ and dashed edges are the edges crossing the partition.
Edge capacities appear below the edges.
Grey vertices denote $U^*=\{v_2,v_3,v_4,v_5\}$ and 
thick edges denote $J^*=\{e_3,e_4\}$.
(b) The corresponding cut in $\mathcal N(T)$.
Grey network vertices are precisely those belonging to
$A^*$, enclosed by the dashed outline.
Arc labels indicate capacities, and all unlabelled
network arcs have capacity $\infty$.
The five bold arcs are exactly those leaving $A^*$.
Here $D^*=1$, and their total capacity is
$\operatorname{cap}_{\mathcal N(T)}(A^*)
=1+\frac12+\frac12+1+1=4=n-1-D^*$.
}
\label{fig:cut-from-partition}
\end{figure}
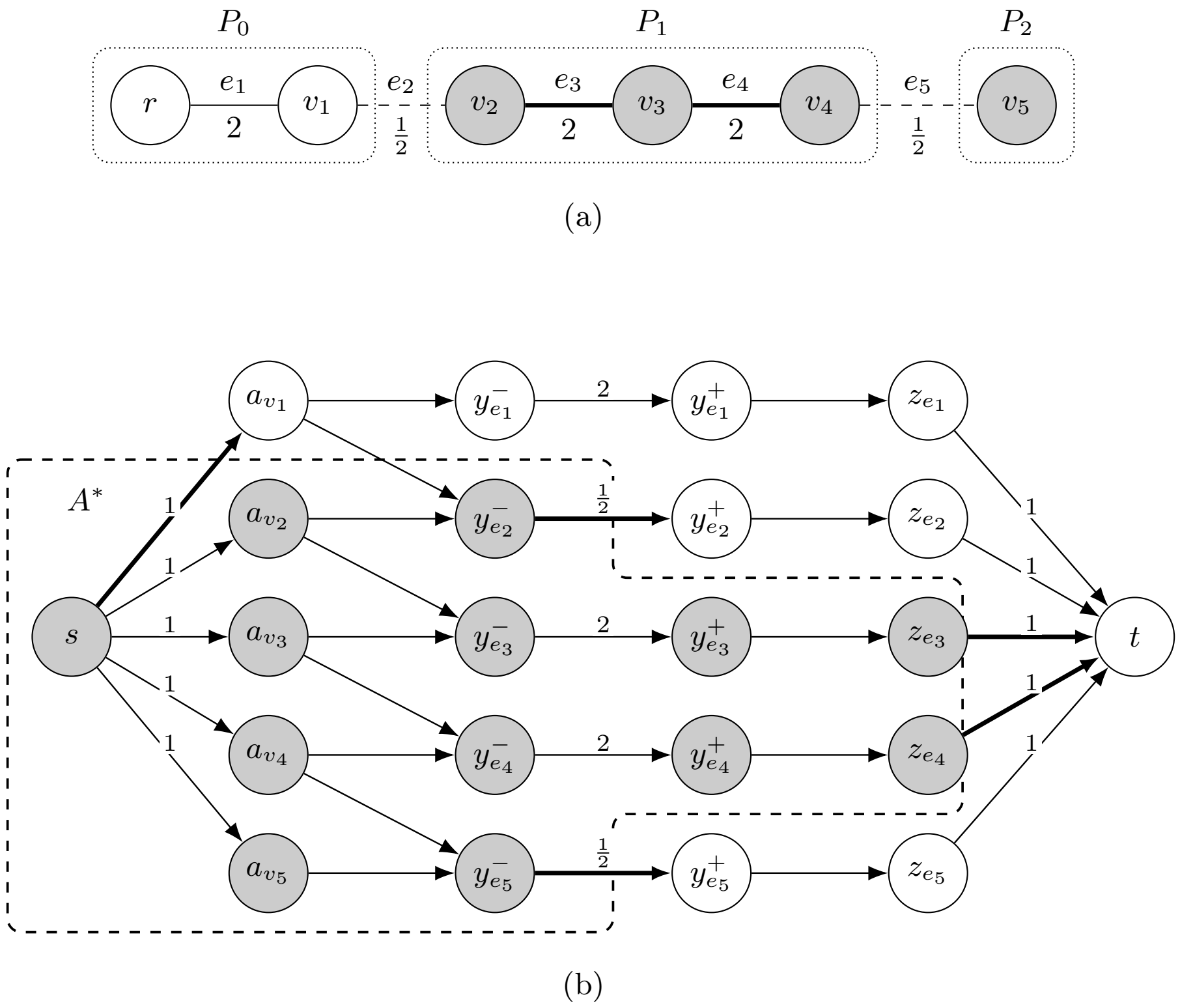

See Figure \ref{fig:cut-from-partition} for an example.
We note that $A^*$ is an $(s,t)$-cut in the network $\mathcal{N}(T)$. We bound the capacity of $A^*$ next. 
We examine the contribution of each kind of arc to the capacity of $A^*$. 
We have a capacity of $n-1-|U^*|$ from arcs leaving the source vertex and a capacity of $|J^*|$ from arcs entering the sink vertex. 
None of the infinite capacity arcs leave $A^*$. 
Among arcs of the form $y_e^-\to y_e^+$, we have a capacity of $q(\delta_G(\Pi^*))$: if both end-vertices of $e$ belong to the same part, then both $y_e^-$ and $y_e^+$ are in $A^*$ and hence they do not contribute and the only contribution is from edges $e$ which have exactly one end-vertex in $A^*$. 
Consequently,
\begin{align*}
 \operatorname{cap}_{\mathcal N(T)}(A^*)
 &=n-1-|U^*|+|J^*|+q(\delta_G(\Pi^*))\\
 &=n-1-(k-1)+q(\delta_G(\Pi^*))\\
 &=n-1-D^*.
\end{align*}
Since $f_T$ is the minimum $(s,t)$-cut capacity, $f_T\le n-1-D^*$ and hence, $\alpha_T\ge D^*$. 
\end{proof}

Next, we show that a tree respecting an optimum partition allows us to recover an optimum partition using one minimum $(s, t)$-cut computation. For a given minimum $(s,t)$-cut in the network $\mathcal{N}(T)$, we associate a partition with it as follows.

\begin{definition}
\label{def:pps-cut-partition}
Let $A$ be the source side of an $(s,t)$-cut in $\mathcal N(T)$. 
Define
\begin{align*}
 U_A &: = \{v\in V\setminus\{r\}:a_v\in A\},\\
 J_A &: = \{f\in E(T):z_f\in A\}.
\end{align*}
Consider the forest $(V, J_A)$ and let $C_1,\ldots,C_b$ be the connected components of this forest whose vertex sets are wholly contained in $U_A$, and let $C_0:=V\setminus \cup_{i=1}^b C_i$. 
The partition associated with $A$ is
\[
 \Pi_A:=\{C_0,C_1,\ldots,C_b\}.
\]
\end{definition}

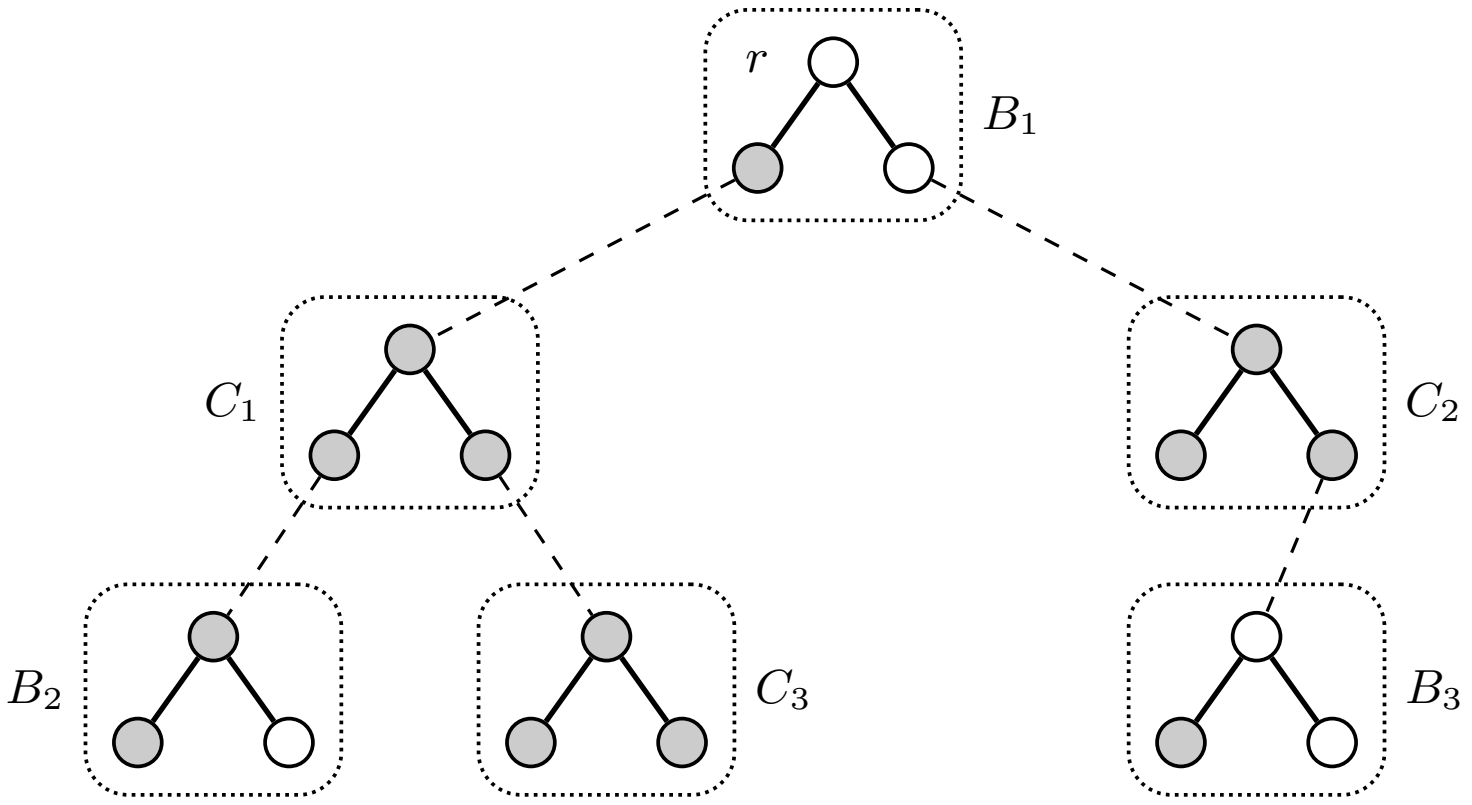
\begin{figure}[t]
\centering
\begin{tikzpicture}[
    x=1cm,y=1cm,
    font=\small,
    vertex/.style={
        circle,draw=black,line width=0.7pt,
        inner sep=0pt,minimum size=9pt
    },
    inside/.style={vertex,fill=black!20,text=black},
    outside/.style={vertex,fill=white},
    jedge/.style={line width=1pt},
    otheredge/.style={dashed,line width=0.6pt},
    component/.style={
        draw,densely dotted,line width=0.7pt,
        rounded corners=8pt,inner sep=5pt
    }
]

\foreach \name/\x/\y in {
    b1l/-0.5/3.8,
    c1/-2.8/2.6,
    c1l/-3.3/1.9,
    c1r/-2.3/1.9,
    c2/2.8/2.6,
    c2l/2.3/1.9,
    c2r/3.3/1.9,
    b2/-4.1/0.7,
    b2l/-4.6/0,
    c3/-1.5/0.7,
    c3l/-2/0,
    c3r/-1/0,
    b3l/2.3/0
}{
    \node[inside] (\name) at (\x,\y) {};
}

\foreach \name/\x/\y in {
    r/0/4.5,
    b1r/0.5/3.8,
    b2r/-3.6/0,
    b3/2.8/0.7,
    b3r/3.3/0
}{
    \node[outside] (\name) at (\x,\y) {};
}

\node[anchor=east] at (-0.3,4.5) {$r$};

\draw[jedge] (b1l)--(r)--(b1r);
\draw[jedge] (c1l)--(c1)--(c1r);
\draw[jedge] (c2l)--(c2)--(c2r);
\draw[jedge] (b2l)--(b2)--(b2r);
\draw[jedge] (c3l)--(c3)--(c3r);
\draw[jedge] (b3l)--(b3)--(b3r);

\draw[otheredge] (b1l)--(c1);
\draw[otheredge] (b1r)--(c2);
\draw[otheredge] (c1l)--(b2);
\draw[otheredge] (c1r)--(c3);
\draw[otheredge] (c2r)--(b3);

\node[component,fit=(r)(b1l)(b1r),
      label=right:{$B_1$}] {};
\node[component,fit=(c1)(c1l)(c1r),
      label=left:{$C_1$}] {};
\node[component,fit=(c2)(c2l)(c2r),
      label=right:{$C_2$}] {};
\node[component,fit=(b2)(b2l)(b2r),
      label=left:{$B_2$}] {};
\node[component,fit=(c3)(c3l)(c3r),
      label=right:{$C_3$}] {};
\node[component,fit=(b3)(b3l)(b3r),
      label=right:{$B_3$}] {};

\end{tikzpicture}
\caption{
The tree $T$ and the partition $\Pi_A$ associated with a $(s,t)$-cut $A$ as in
Definition~\ref{def:pps-cut-partition}.
Solid lines indicate edges of $J_A$, and dashed lines
indicate edges of $E(T)\setminus J_A$.
Grey vertices belong to $U_A$, whereas white vertices
lie outside $U_A$.
Dotted outlines enclose the connected components of $(V,J_A)$.
The components $C_1,C_2,C_3$ are wholly contained in $U_A$
and are retained as separate parts.
Each of $B_1,B_2,B_3$ contains a vertex outside $U_A$,
so these components are merged into
$C_0=B_1\cup B_2\cup B_3$.
Thus $\Pi_A=\{C_0,C_1,C_2,C_3\}$.
We note that $T[C_0]$ is disconnected.
}
\label{fig:pps-cut-partition}
\end{figure}

See Figure \ref{fig:pps-cut-partition} for an example. Let $A$ be an $(s, t)$-cut in $\mathcal{N}(T)$ and let $\Pi_A=\{C_0, C_1, C_2, \ldots, C_b\}$ be the partition associated with $A$ as defined in Definition \ref{def:pps-cut-partition}. The root vertex $r$ is not in $U_A$. Therefore, $C_0$ contains $r$ and is non-empty, and $|\Pi_A|-1=b$. Our second lemma shows that the shortfall is at most the deficit of the partition recovered from a minimum $(s,t)$-cut of $\mathcal{N}(T)$. While Lemma \ref{lem:pps-respecting-cut} holds only if $T$ is $\Pi^*$ respecting, the next lemma holds even if $T$ is not $\Pi^{*}$-respecting.

\begin{lemma}
\label{lem:pps-cut-deficit}
Let $T$ be an arbitrary tree on $V$, $A$ be the
source side of a minimum $(s,t)$-cut in $\mathcal N(T)$, and $\Pi_A$ be the partition associated with $A$ as in Definition \ref{def:pps-cut-partition}.
Then,
\[
    \alpha_T \le D(\Pi_A)\le D^*. 
\]
\end{lemma}

\begin{proof}
Let $U_A,J_A,C_0,C_1,\ldots,C_b$ be as in 
Definition~\ref{def:pps-cut-partition}, and let $\mathcal C$
be the connected components of the forest $(V,J_A)$.
Since each component $C$ of $(V, J_A)$ contains $|C|-1$ edges of $J_A$, we have that 
\[
    |U_A|-|J_A|
    =\sum_{C\in\mathcal C}
       \bigl(|U_A\cap C|-|J_A\cap E[C]|\bigr)
    =\sum_{C\in\mathcal C}
       \bigl(1-|C\setminus U_A|\bigr)
    \le |\Pi_A|-1.
\]
The last inequality is because, each component $C\in \mathcal{C}$ wholly contained in $U_A$ contributes
one to the sum, whereas every other component $C\in \mathcal{C}$ contributes at most zero to the sum and in particular, the component containing the root vertex $r$ contributes zero to the sum. 

Next, we lower bound the cut capacity of $A$ as a function of the partition $\Pi_A$. 
We recall that $\mathcal{N}(T)$ has a trivial $(s,t)$-cut of capacity $n-1$, hence the capacity of the minimum $(s,t)$-cut $A$ is $f_T\le n-1$. Thus, no infinite-capacity arc leaves $A$. 
Consider an edge $e=uv\in\delta_G(\Pi_A)$.
At least one end-vertex, say $u$, belongs to a 
component $C_i$ for some $i\in [b]$, and hence $u\in U_A$.
Thus, $a_u\in A$, and the infinite capacity arc $a_u\to y_e^-$ forces
$y_e^-\in A$.
The end-vertices of $e$ lie in distinct components of
the forest $(V,J_A)$, so their tree path $P_T(u,v)$ contains an edge
$f\notin J_A$.
Consequently, $z_f\notin A$, and the infinite capacity arc
$y_e^+\to z_f$ forces $y_e^+\notin A$.
Therefore, the arc $y_e^-\to y_e^+$, of capacity
$q_e$, leaves $A$.
The outgoing arcs from the source leaving $A$ contribute $n-1-|U_A|$,
and the incoming arcs to the sink leaving $A$ contribute $|J_A|$ to the capacity of $A$.
another and from these source and sink arcs.
Consequently, the capacity $f_T$ of the $(s,t)$-cut $A$ satisfies 
\[
    f_T\ge n-1-|U_A|+|J_A|+q(\delta_G(\Pi_A))\ge n-|\Pi_A|+q(\delta_G(\Pi_A)), 
\]
where the last inequality is using $|U_A|-|J_A|\le b=|\Pi_A|-1$. Thus, we obtain
\begin{align*}
    \alpha_T
    &\le |\Pi_A|-1-q(\delta_G(\Pi_A))
     =D(\Pi_A)
     \le D^*.
\end{align*}
The last inequality follows from the definition of $D^*$.
\end{proof}

We now prove Lemma~\ref{lem:pps-recover-optimum}. In particular, we show that if $T$ is a $\Pi^*$ respecting tree for some optimal partition $\Pi^*$ to the Lagrangian problem on $G$ and $A$ is a minimum $(s,t)$-cut in $\mathcal{N}(T)$, then $\Pi_{A}$ is also an optimal partition and can be recovered in $O(m+n)$ time. The run-time claimed in Lemma~\ref{lem:pps-recover-optimum} is obtained by working with $\compact(T)$ instead of $\mathcal{N}(T)$. 

\begin{proof}[Proof of Lemma~\ref{lem:pps-recover-optimum}]
We first describe the algorithm. Construct $\compact(T)$
using Lemma~\ref{lem:pps-compact-network}, implementing
every infinite-capacity arc with capacity $n$, and compute
a maximum $(s,t)$-flow in this network. Obtain the source
side $A_{\mathrm{compact}}$ of a minimum $(s,t)$-cut by a
reachability search from $s$ in the residual network.
Set
\[
    A:=A_{\mathrm{compact}}\cap V(\mathcal N(T)),
\]
construct $\Pi_A$ as in
Definition~\ref{def:pps-cut-partition}, and return $\Pi_A$.

Next, we prove correctness. The arcs leaving $s$ have total
capacity $n-1$, so $f_T\le n-1$. Replacing infinite capacities
by $n$ preserves the minimum cuts and their capacities:
the source cut has capacity $n-1$, whereas any cut crossing
a replaced arc has capacity at least $n$. By the
flow-preservation property of
Lemma~\ref{lem:pps-compact-network}, the computed maximum
flow has value $f_T$. Consequently, $A_{\mathrm{compact}}$
has capacity $f_T\le n-1$, and the cut-preservation property
of the same lemma implies that $A$ is the source side of
a minimum $(s,t)$-cut in $\mathcal N(T)$ with capacity $f_T$.
By Lemmas~\ref{lem:pps-respecting-cut}
and~\ref{lem:pps-cut-deficit},
\[
    D^*\le \alpha_T\le D(\Pi_A)\le D^*.
\]
Thus all inequalities are equalities, and $\Pi_A$ is a
deficit-maximizing partition and hence an optimal partition
for the Lagrangian problem on $G$.

Finally, we prove the running-time guarantees.
By Lemma~\ref{lem:pps-compact-network}, the compact network
has $O(m+n)$ vertices and $O(n+m\log^2 n)$ arcs, and its
construction takes $O(n+m\log^2 n)$ arithmetic operations.
The retained arcs have capacities $1$, $q_e$, or $n$,
and every auxiliary arc introduced by the compact
construction has capacity $n$. Thus every arc capacity
belongs to
\[
    \{1,n\}\cup\{q_e:e\in E\}.
\]
Constructing the residual network and performing the
reachability search take a number of arithmetic operations
linear in the compact network size. Restricting
$A_{\mathrm{compact}}$ to the original network nodes takes
$O(m+n)$ time and does not require explicitly constructing
$\mathcal N(T)$. To construct $\Pi_A$, read $U_A$ and $J_A$
from $A$, and traverse the forest $(V,J_A)$ to identify its
components. Retain each component wholly contained in $U_A$
as a separate part, and merge the remaining components
into $C_0$, as in
Definition~\ref{def:pps-cut-partition}.
These steps take $O(m+n)$ time. This proves the bound of
one maximum-flow computation together with
$O((m+n)\log^2 n)$ additional arithmetic operations.

For the second guarantee, fix $S\ge m+n$ and a constant
$C>0$. Suppose that $q_e=a_e/B$ for every $e\in E$,
where $B$ and all $a_e$ are positive integers at most $S^C$,
and that the returned flow values are integer multiples
of $1/B$. Represent capacities and flow values using
their integer numerators over the common denominator $B$.
The capacity numerators belong to
\[
    \{B,nB\}\cup\{a_e:e\in E\}.
\]
Since $n\le S$, every such numerator is at most $S^{C+1}$.
Feasibility bounds each flow numerator by the corresponding
capacity numerator. Hence the capacity numerators, flow
numerators, and numerators of residual capacities all have
$O(\log S)$ bits, with the implied constant depending only
on $C$. Residual capacities and their positivity can
therefore be computed using arithmetic on integers of
this bit length. All remaining operations manipulate graph
indices and counters whose bit lengths are also
$O(\log S)$. Consequently, each additional arithmetic
operation can be implemented in $\log^{O(1)}S$ time.
Since $\log n\le\log S$, the total running time excluding
the maximum-flow computation is
\[
    O\!\left((m+n)\log^2 n\log^{O(1)}S\right)
    =
    O\!\left((m+n)\log^{O(1)}S\right),
\]
where the exponent depends only on $C$.

\end{proof}

\section{Finding a respecting tree}
\label{sec:find-respecting-tree}

Let $G=(V,E,q:E\to\mathbb{Q}_{>0})$ be an undirected graph with positive rational capacities and a fixed root vertex $r\in V$, where $n:=|V|$ and $m:=|E|$. In this section, we show that a $\Pi^*$-respecting tree can be found using logarithmic max-flows in networks of near-linear size, where $\Pi^*$ is an optimum partition for the Lagrangian problem on $G$.

\begin{lemma}
\label{thm:pps-find-respecting-tree}
Given an undirected graph
$G=(V,E,q:E\to\mathbb{Q}_{>0})$ with $n\ge1$ vertices
and $m$ edges, and $\eta\in(0,1)$, there is a randomized
algorithm that returns a $\Pi^*$-respecting tree with
probability at least $1-\eta$, for every fixed optimum
partition $\Pi^*$ to the Lagrangian problem on $G$.
For $n=1$, the algorithm takes constant time and requires
no maximum-flow computation. For $n\ge2$, it has the
following running-time guarantees:
\begin{enumerate}
\item The algorithm performs $O(\log(n/\eta))$ iterations,
each using one maximum-flow computation on a network with
$O(m+n)$ vertices and $O(n+m\log^2 n)$ arcs, where every
arc capacity belongs to $\{1,n\}\cup\{q_e:e\in E\}$,
together with
\[
    O\!\left((n+m\log^2 n)\log(m+n)\right)
\]
additional expected arithmetic operations per iteration.

\item Let $S\ge m+n$ and let $C>0$ be fixed.
Suppose that $q_e=p_e/B$ for every $e\in E$, where $B$
and all $p_e$ are positive integers at most $S^C$,
and suppose that $1/\eta\le S^C$.
If every maximum-flow computation returns a flow whose
arc values are integer multiples of $1/B$, then the total
expected running time excluding the maximum-flow
computations is
\[
    O\!\left((m+n)\log^{O(1)}S\right),
\]
where the exponent depends only on $C$.
\end{enumerate}
\end{lemma}

Throughout this section, we fix an optimum partition $\Pi^*$ for the Lagrangian problem on $G$. This partition is fixed only for the purpose of the analysis and is not known to the algorithm. 
Let $T$ be an arbitrary tree on $V$. 
We define the \emph{$\Pi^*$-excess} of $T$ by
\[
 z_{\Pi^*}(T)
 :=|\delta_T(\Pi^*)|-(|\Pi^*|-1).
\]
Every tree has at least $|\Pi^*|-1$ edges crossing $\Pi^*$.
Thus, $z_{\Pi^*}(T)$ is a nonnegative integer, and it is zero
exactly when $T$ respects $\Pi^*$.

Our key ingredient to prove Lemma \ref{thm:pps-find-respecting-tree} is a randomized algorithm that takes an arbitrary tree $T$ and produces a tree $T'$ whose $\Pi^*$-excess is at most half that of $T$ in expectation. 

\begin{lemma}
\label{lem:halving}
Given an undirected graph
$G=(V,E,q:E\rightarrow\mathbb{Q}_+)$ with $n\ge1$ vertices
and $m$ edges, and a tree $T$ on $V$, there is a randomized
algorithm that returns a tree $T'$ on $V$ such that, for
every fixed optimum partition $\Pi^*$ to the Lagrangian
problem on $G$,
\begin{equation}
\label{eq:pps-update-halving}
    \mathbb E\!\left[z_{\Pi^*}(T')\right]
    \le \frac12 z_{\Pi^*}(T).
\end{equation}
For $n=1$, the algorithm takes constant time and requires
no maximum-flow computation. For $n\ge2$, it has the
following running-time guarantees:
\begin{enumerate}
\item The algorithm uses one maximum-flow computation on
a network with $O(m+n)$ vertices and $O(n+m\log^2 n)$ arcs,
where every arc capacity belongs to
$\{1,n\}\cup\{q_e:e\in E\}$, together with
\[
    O\!\left((n+m\log^2 n)\log(m+n)\right)
\]
additional expected arithmetic operations.

\item Let $S\ge m+n$ and let $C>0$ be fixed.
Suppose that $q_e=a_e/B$ for every $e\in E$, where $B$
and all $a_e$ are positive integers satisfying
$B\le S^C$ and $a_e\le S^C$ for every $e\in E$. 
If the maximum-flow computation returns a flow whose arc
values are integer multiples of $1/B$, then the additional
expected running time is
\[
    O\!\left((m+n)\log^{O(1)}S\right),
\]
where the exponent depends only on $C$.
\end{enumerate}
\end{lemma}

We defer the proof of Lemma \ref{lem:halving} and first use it to prove Lemma \ref{thm:pps-find-respecting-tree}. 

\begin{proof}[Proof of Lemma~\ref{thm:pps-find-respecting-tree}]
We first describe the algorithm. If $n=1$, return the tree
with no edges. Henceforth, assume that $n\ge2$.
Choose an arbitrary tree $T_0$ on $V$ and set
\[
    R:=\left\lceil\log_2((n-1)/\eta)\right\rceil.
\]
For every $i\in\{0,1,\ldots,R-1\}$, apply the algorithm
of Lemma~\ref{lem:halving} to $G$ and $T_i$, using fresh
randomness, to obtain a tree $T_{i+1}$.
Return $T_R$.

Next, we prove correctness. The case $n=1$ is immediate,
so suppose that $n\ge2$. Fix an optimum partition $\Pi^*$
to the Lagrangian problem on $G$, independently of the
algorithm's random choices.
By Lemma~\ref{lem:halving} and the use of fresh randomness,
for every $i\in\{0,1,\ldots,R-1\}$,
\[
    \mathbb E\!\left[z_{\Pi^*}(T_{i+1})\mid T_i\right]
    \le \frac12 z_{\Pi^*}(T_i).
\]
Taking expectations and iterating this inequality gives
\[
    \mathbb E\!\left[z_{\Pi^*}(T_R)\right]
    \le 2^{-R}z_{\Pi^*}(T_0)
    \le 2^{-R}(n-1)
    \le \eta.
\]
Since the excess is a nonnegative integer and is zero
exactly when the tree respects $\Pi^*$, Markov's
inequality yields
\[
\begin{aligned}
    \Pr[T_R\text{ does not respect }\Pi^*]
    &=\Pr[z_{\Pi^*}(T_R)\ge1]\\
    &\le \mathbb E[z_{\Pi^*}(T_R)]\\
    &\le \eta.
\end{aligned}
\]
Thus the returned tree respects $\Pi^*$ with probability
at least $1-\eta$. Since $\Pi^*$ was an arbitrary fixed
optimum partition, this proves the claimed guarantee.

Finally, we prove the running-time guarantees.
The case $n=1$ takes constant time and requires no
maximum-flow computation. Suppose that $n\ge2$.
There are $R=O(\log(n/\eta))$ iterations.
By the first guarantee of Lemma~\ref{lem:halving},
each iteration uses one maximum-flow computation on
a network with $O(m+n)$ vertices and $O(n+m\log^2 n)$ arcs,
where every arc capacity belongs to
\[
    \{1,n\}\cup\{q_e:e\in E\},
\]
together with
\[
    O\!\left((n+m\log^2 n)\log(m+n)\right)
\]
additional expected arithmetic operations.
Summing over the iterations gives
\[
    O\!\left(
        R(n+m\log^2 n)\log(m+n)
    \right)
\]
additional expected arithmetic operations.
The $O(n)$ time to construct $T_0$ is absorbed by this
bound.

For the second guarantee, fix $S\ge m+n$ and a constant
$C>0$. Suppose that $q_e=p_e/B$ for every $e\in E$,
where $B$ and all $p_e$ are positive integers at most
$S^C$, that $1/\eta\le S^C$, and that every maximum-flow
computation returns a flow whose arc values are integer
multiples of $1/B$.
The graph and its capacities remain unchanged throughout
the algorithm; only the tree changes.
Consequently, for every realized input tree $T_i$,
the second guarantee of Lemma~\ref{lem:halving} applies
with the same $S$ and $C$. It bounds the conditional
expected running time outside the maximum-flow computation
in that iteration by
\[
    O\!\left((m+n)\log^{O(1)}S\right),
\]
uniformly over the choice of $T_i$.
Moreover, $n\le S$ and $1/\eta\le S^C$ imply
\[
    R
    \le 1+\log_2(n-1)+\log_2(1/\eta)
    \le 1+(C+1)\log_2 S.
\]
By the law of total expectation and linearity of
expectation, the total expected running time excluding
the maximum-flow computations is therefore
\[
\begin{aligned}
    O(n)+R\,O\!\left((m+n)\log^{O(1)}S\right)
    &=O\!\left((m+n)\log^{O(1)}S\right),
\end{aligned}
\]
where the exponent depends only on $C$.
\end{proof}

The remainder of this section proves Lemma~\ref{lem:halving}. 
We give the algorithm and prove correctness in
Section~\ref{sec:halving-algo}, and show an efficient implementation in
Section~\ref{sec:halving-runtime}.
Let $G=(V, E, c: E\rightarrow \mathbb{Q}_+)$ be the input graph and $T$ be the input tree for both these sections. We will use the following standard randomized flow-rounding result---e.g., see Sections 3.3 and 4 of \cite{KP15}.

\begin{lemma}\cite{KP15}
\label{lem:flow-rounding}
Let $g$ be a nonnegative rational $(s,t)$-flow of
integer value $L$ on a directed network. There is a randomized
algorithm that returns an integral
$(s,t)$-flow $g_{\mathrm{int}}$ of value $L$ on the
same arcs such that, for every arc $a$,
\begin{subequations}\label{eq:rounding-marginals}
\begin{align}
 \lfloor g(a)\rfloor
 \le g_{\mathrm{int}}(a)
 &\le\lceil g(a)\rceil, \text{ and}
 \label{eq:rounding-bounds}\\
 \mathbb E[g_{\mathrm{int}}(a)]
 &=g(a).
 \label{eq:rounding-expectation}
\end{align}
\end{subequations}
\end{lemma}

\subsection{Algorithm and Correctness}\label{sec:halving-algo}
In this section, we describe an algorithm to prove Lemma \ref{lem:halving} and prove its correctness. We describe the algorithm using the network $\mathcal{N}(T)$ for simplicity. Efficient implementation to achieve the desired run-time will be done using the compact representation $\compact(T)$ from Section \ref{sec:flow}---these implementation details will be discussed in the next subsection. 

We give an outline of the algorithm now. We begin with a rational maximum flow $f$ in the network $\mathcal{N}(T)$. We recall that $f_T\le n-1$. The flow $f$ might saturate some of the arcs leaving the source $s$ and some of the arcs entering the sink $t$. 
We augment the network to $\augment(T)$ by adding one dummy node $h$, 
unit-capacity arcs from all vertices $a_v$ to the node $h$ (exactly $n-1$ such arcs), and 
unit-capacity arcs from $h$ to all vertices $z_f$ (exactly $n-1$ such arcs). We augment the flow $f$ along these newly added arcs to obtain flow $f^{\mathrm{aug}}$ in $\augment(T)$ such that the value of $f^{\mathrm{aug}}$ is exactly $n-1$ with the amount of the flow being routed through $h$ being exactly the shortfall $\alpha_T$. Next, we use Lemma \ref{lem:flow-rounding} to randomly round $f^{\mathrm{aug}}$ to obtain an integral flow $f^{\mathrm{aug}}_{int}$ in $\augment(T)$. We use the integral flow $f^{\mathrm{aug}}_{int}$ to obtain a multi-set $F$ of edges such that the multi-graph $(V, F\uplus E(T))$ has $2$ edge-disjoint spanning trees. We find two edge-disjoint spanning trees in $(V, F\uplus E(T))$, say $T_1$ and $T_2$, sample $T'\in \{T_1, T_2\}$ uniformly at random and return it. 
In the rest of the section, we formalize the above-described approach and prove correctness.

\paragraph{Augmenting the network and the max flow.} We begin with the description of the augmented network and the augmented flow. 

\begin{definition}\label{def:augment}
Let $f$  be a maximum flow in $\mathcal{N}(T)$ with value $f_T$. Let $\augment(T)$ be obtained from $\mathcal{N}(T)$ by adding a dummy
vertex $h$, unit-capacity arcs $a_v\to h$ for all
$v\in V\setminus\{r\}$, and unit-capacity arcs $h\to z_g$ for all
$g\in E(T)$. For all $v\in V\setminus\{r\}$, define
\begin{align*}
 \augflow(s,a_v)&=1, \text{ and}\\
 \augflow(a_v,h)&=1-f(s,a_v).
\end{align*}
For all $g\in E(T)$, define
\begin{align*}
 \augflow(z_g,t)&=1, \text{ and}\\
 \augflow(h,z_g)&=1-f(z_g,t).
\end{align*}
For every other arc $(x,y)$, let $\augflow(x,y)=f(x,y)$.
\end{definition}

\begin{figure}[t]
\centering
\begin{tikzpicture}[
  x=1cm,y=1cm,
  font=\small,
  vertex/.style={
    circle,draw=black,fill=gray!12,text=black,
    line width=0.6pt,minimum size=9mm,inner sep=1pt
  },
  arc/.style={
    -{Stealth[length=1.8mm]},line width=0.6pt
  },
  increased/.style={
    arc,line width=1.25pt
  },
  added/.style={
    arc,dashed,
    preaction={draw=white,line width=3pt}
  },
  addedpositive/.style={
    added,line width=1.25pt
  },
  flowlabel/.style={
    fill=white,inner sep=1.5pt,text=black
  }
]

\begin{scope}[yshift=4.4cm]
  \node[vertex] (os) at (0,0) {$s$};
  \node[vertex] (oau) at (2.2,0.9) {$a_u$};
  \node[vertex] (oav) at (2.2,-0.9) {$a_v$};
  \node[vertex] (om1) at (4.5,0.9) {$y^-_{e_1}$};
  \node[vertex] (om2) at (4.5,-0.9) {$y^-_{e_2}$};
  \node[vertex] (op1) at (6.7,0.9) {$y^+_{e_1}$};
  \node[vertex] (op2) at (6.7,-0.9) {$y^+_{e_2}$};
  \node[vertex] (oz1) at (9,0.9) {$z_{e_1}$};
  \node[vertex] (oz2) at (9,-0.9) {$z_{e_2}$};
  \node[vertex] (ot) at (11.2,0) {$t$};

  \draw[arc] (os)
    -- node[flowlabel,above] {$1$} (oau);
  \draw[arc] (os)
    -- node[flowlabel,below] {$\frac12$} (oav);

  \draw[arc] (oau)
    -- node[flowlabel,above] {$\frac12$} (om1);
  \draw[arc] (oau)
    -- node[flowlabel,above right] {$\frac12$} (om2);
  \draw[arc] (oav)
    -- node[flowlabel,below] {$\frac12$} (om2);

  \draw[arc] (om1)
    -- node[flowlabel,above] {$\frac12$} (op1);
  \draw[arc] (om2)
    -- node[flowlabel,below] {$1$} (op2);

  \draw[arc] (op1)
    -- node[flowlabel,above] {$\frac12$} (oz1);
  \draw[arc] (op2)
    -- node[flowlabel,below] {$1$} (oz2);

  \draw[arc] (oz1)
    -- node[flowlabel,above] {$\frac12$} (ot);
  \draw[arc] (oz2)
    -- node[flowlabel,below] {$1$} (ot);

  \node at (5.6,-1.8)
    {(a) Maximum flow $f$ in $\mathcal N(T)$,
     of value $3/2$.};
\end{scope}

\node[vertex] (cs) at (0,0) {$s$};
\node[vertex] (cau) at (2.2,0.9) {$a_u$};
\node[vertex] (cav) at (2.2,-0.9) {$a_v$};
\node[vertex] (cm1) at (4.5,0.9) {$y^-_{e_1}$};
\node[vertex] (cm2) at (4.5,-0.9) {$y^-_{e_2}$};
\node[vertex] (cp1) at (6.7,0.9) {$y^+_{e_1}$};
\node[vertex] (cp2) at (6.7,-0.9) {$y^+_{e_2}$};
\node[vertex] (cz1) at (9,0.9) {$z_{e_1}$};
\node[vertex] (cz2) at (9,-0.9) {$z_{e_2}$};
\node[vertex] (ct) at (11.2,0) {$t$};
\node[vertex,fill=gray!25] (h) at (5.6,-2.9) {$h$};

\draw[arc] (cs)
  -- node[flowlabel,above] {$1$} (cau);
\draw[increased] (cs)
  -- node[flowlabel,below] {$1$} (cav);

\draw[arc] (cau)
  -- node[flowlabel,above] {$\frac12$} (cm1);
\draw[arc] (cau)
  -- node[flowlabel,above right] {$\frac12$} (cm2);
\draw[arc] (cav)
  -- node[flowlabel,below,pos=0.68] {$\frac12$} (cm2);

\draw[arc] (cm1)
  -- node[flowlabel,above] {$\frac12$} (cp1);
\draw[arc] (cm2)
  -- node[flowlabel,below] {$1$} (cp2);

\draw[arc] (cp1)
  -- node[flowlabel,above] {$\frac12$} (cz1);
\draw[arc] (cp2)
  -- node[flowlabel,below,pos=0.76] {$1$} (cz2);

\draw[increased] (cz1)
  -- node[flowlabel,above] {$1$} (ct);
\draw[arc] (cz2)
  -- node[flowlabel,below] {$1$} (ct);

\draw[added] (cau.south)
  .. controls (3.1,-0.2) and (3.0,-2.1) ..
  node[flowlabel,pos=0.67,above right] {$0$}
  (h.north west);

\draw[addedpositive] (cav.south)
  .. controls (2.2,-2.9) and (3.8,-2.9) ..
  node[flowlabel,pos=0.57,below] {$\frac12$}
  (h.west);

\draw[addedpositive] (h.north east)
  .. controls (7.2,-1.8) and (7.8,0) ..
  node[flowlabel,pos=0.28,below right] {$\frac12$}
  (cz1.south west);

\draw[added] (h.east)
  .. controls (7.5,-2.9) and (9,-2.4) ..
  node[flowlabel,pos=0.53,below] {$0$}
  (cz2.south);

\node at (5.6,-3.8)
  {(b) Completed flow $f^{\mathrm{aug}}$ in
   $\mathcal N_f^{\mathrm{aug}}(T)$, of value $2$.};

\end{tikzpicture}

\caption{Completing a maximum flow. Here $G$ is given by the path
$r-u-v$, and the spanning tree $T$ is the entire graph itself. Let $e_1=ru$, $e_2=uv$, $q_{e_1}=1/2$,
and $q_{e_2}=1$. Arc labels give flow values, not capacities.
In (a), the flow saturates both shared arcs and is therefore
maximum. In (b), dashed arcs are the new unit-capacity arcs
incident with $h$. The thick path carries the additional
half-unit along $s\to a_v\to h\to z_{e_1}\to t$.
Every source and sink arc then carries one unit, while all
other old arc flows remain unchanged. The completed flow
has value $2=n-1$, and the flow through $h$ is
$1/2=\alpha_T$.}
\label{fig:augment}
\end{figure}
See Figure \ref{fig:augment} for an example. We summarize the key properties of the augmented flow below. 
\begin{proposition}\label{prop:augmented-flow}
    Let $f$ be a maximum flow in $\mathcal N(T)$ with value $f_T$.
Let $\augment(T)$ and $\augflow$ be as in
Definition~\ref{def:augment}. Then, the following hold: 
\begin{enumerate}
\renewcommand{\theenumi}{\roman{enumi}}
\renewcommand{\labelenumi}{(\theenumi)}
 \item\label{prop:augmented-flow:value}
 The flow $\augflow$ is feasible in $\augment(T)$ and has value $n-1$.
 \item\label{prop:augmented-flow:edge}
 For all $e\in E$, we have that 
 $\augflow(y_e^-,y_e^+)=f(y_e^-,y_e^+)\le q_e$. 
 \item\label{prop:augmented-flow:dummy}
 The total flow through $h$ is $\alpha_T$, that is, $\sum_{v\in V\setminus\{r\}}\augflow(a_v,h)=\alpha_T$.
\end{enumerate}
\end{proposition}
\begin{proof}
    For every $v\in V\setminus \{r\}$, the additional incoming flow at $a_v$ is $1-f(s,a_v)$, which is exactly the flow sent on $a_v\to h$. For every $g\in E(T)$, the additional outgoing flow from $z_g$ is $1-f(z_g, t)$, which is supplied by $h\to z_g$. At $h$, we have 
    \begin{align*}
 \sum_{v\in V\setminus\{r\}}\augflow(a_v,h)
 &=\sum_{v\in V\setminus\{r\}}(1-f(s,a_v))
 =n-1-f_T=\alpha_T,\\
 \sum_{g\in E(T)}\augflow(h,z_g)
 &=\sum_{g\in E(T)}(1-f(z_g,t))
 =n-1-f_T=\alpha_T.
\end{align*}
These equalities prove conservation at $h$ and
part~(\ref{prop:augmented-flow:dummy}). Every new arc carries a
value in $[0,1]$, and all other capacity constraints and conservation
equations remain satisfied. Every arc leaving $s$ carries one unit,
so the value of the flow is $n-1$, proving part~(\ref{prop:augmented-flow:value}).
For every $e\in E$, the arc $y_e^-\to y_e^+$ retains its original
flow, proving part~(\ref{prop:augmented-flow:edge}).
\end{proof}

\paragraph{Rounding the Augmented Flow.} Applying Lemma \ref{lem:flow-rounding} to the augmented flow gives the following guarantees. 

\begin{proposition}\label{prop:rounded-flow}
    Let $f$ be a rational maximum flow in $\mathcal{N}(T)$ with value $f_T$. 
    Let $\augment(T)$ and $\augflow$ be as in Definition \ref{def:augment}, and let $\intaugflow$ be obtained by applying Lemma \ref{lem:flow-rounding} to $\augflow$. Then the following hold: 
    \begin{enumerate}
        \item\label{prop:rounded-flow:value} 
        The flow $\intaugflow$ is integral and has value $n-1$. Moreover, for all $v\in V\setminus \{r\}$, we have $\intaugflow(s, a_v)=1$ and for all $g\in E(T)$, we have $\intaugflow(z_g, t)=1$. 
        \item\label{prop:rounded-flow:edge} 
        For every graph edge $e\in E$, we have $\E[\intaugflow(y_e^-, y_e^+)]\le q_e$. 
        \item\label{prop:rounded-flow:dummy} 
        The expected total flow through $h$ satisfies $\E[\sum_{v\in V\setminus \{r\}}\intaugflow(a_v,h)]=\alpha_T$. 
    \end{enumerate}
\end{proposition}
\begin{proof}
    By Proposition~\ref{prop:augmented-flow}(\ref{prop:augmented-flow:value}),
the augmented flow has value $n-1$.
Lemma~\ref{lem:flow-rounding} preserves this value and returns an
integral flow. The bounds~\eqref{eq:rounding-bounds} preserve the
unit values on every source and sink arc, proving
part~(\ref{prop:rounded-flow:value}). The expectation guarantee
\eqref{eq:rounding-expectation}, together with
Proposition~\ref{prop:augmented-flow}(\ref{prop:augmented-flow:edge}),
gives part~(\ref{prop:rounded-flow:edge}). Finally, linearity of
expectation and~\eqref{eq:rounding-expectation} give
\[
 \mathbb E\!\left[
  \sum_{v\in V\setminus\{r\}}\intaugflow(a_v,h)
 \right]
 =\sum_{v\in V\setminus\{r\}}\augflow(a_v,h).
\]
Proposition~\ref{prop:augmented-flow}(\ref{prop:augmented-flow:dummy})
implies that the last sum is $\alpha_T$, proving
part~(\ref{prop:rounded-flow:dummy}).
\end{proof}

\paragraph{Selecting edges from the rounded flow.} 
We define the multiset $F$ together with a vertex and tree-edge assignments that will be used to show that $E(T)\uplus F$ contains two edge-disjoint spanning trees. 
\begin{definition}\label{def:selected-edges}
Let $f$ be a maximum flow in $\mathcal{N}(T)$ with value $f_T$. 
Let $\augment(T)$ and $f^{\mathrm{aug}}$ be as in
Definition~\ref{def:augment}, and let
$f^{\mathrm{aug}}_{\mathrm{int}}$ be an integral flow obtained by
applying Lemma~\ref{lem:flow-rounding}. Decompose
$f^{\mathrm{aug}}_{\mathrm{int}}$ into $n-1$ unit $(s,t)$-paths.
For each $v\in V\setminus\{r\}$, let $Q_v$ be the unique path starting with
$s\to a_v$, and let $g_v\in E(T)$ be the tree edge whose node
$z_{g_v}$ is the last internal node of $Q_v$. For each $v\in V\setminus \{r\}$, we select a tuple $(e_v, g_v)$, where $e_v=\{v,w_v\}$ is a copy of an edge of $G$ and $g_v$ is an edge of $T$, as follows: 
\begin{enumerate}
 \item If $Q_v$ avoids $h$, it has the following form for some $e\in \delta_G(v)$ and some tree edge $g\in E(T)$: 
 \[
  s\to a_v\to y_e^-\to y_e^+\to z_{g}\to t.
 \]
 Let $w_v$ be the other end-vertex of the edge $e$ incident
 with $v$, and let $e_v$ be a copy of $e$. Let $g_v:=g$. 
 \item If $Q_v$ passes through $h$, it has the following form for some tree edge $g\in E(T)$: 
 \[
  s\to a_v\to h\to z_{g}\to t.
 \]
 Let $w_v$ be the end-vertex of $g_v$ in the component of $T-g_v$
 not containing $v$, and let $e_v$ be a copy of $\{v,w_v\}$. Let $g_v:=g$. 
 Call this copy of $e_v$ a \emph{dummy edge} (since it need not belong to $G$).
\end{enumerate}
Let $F:=\{e_v:v\in V\setminus\{r\}\}$ be the multiset of the
selected copies. Copies remain distinct even when
their end-vertices coincide with those of another copy or an edge
of $T$. %
\end{definition}

We have the following properties about the mapping $v\mapsto (e_v, g_v)$ constructed in Definition \ref{def:selected-edges}. 
\begin{proposition}\label{prop:selected-edges}
The construction in Definition~\ref{def:selected-edges} is
well-defined and has the following properties.
\begin{enumerate}
\renewcommand{\theenumi}{\roman{enumi}}
\renewcommand{\labelenumi}{(\theenumi)}
 \item\label{prop:selected-edges:bijection}
 The assignment $v\mapsto g_v$ is a bijection from
 $V\setminus\{r\}$ to $E(T)$.
 \item\label{prop:selected-edges:path}
 For all $v\in V\setminus\{r\}$,
 \begin{equation}\label{eq:selected-path-property}
  g_v\in P_T(v,w_v).
 \end{equation}
\end{enumerate}
\end{proposition}
\begin{proof}
By Proposition~\ref{prop:rounded-flow}(\ref{prop:rounded-flow:value}),
the rounded flow is integral, has value $n-1$, and carries one
unit on every outgoing arc from the source and incoming arc into the sink. Since $\augment(T)$ is acyclic,
this flow decomposes into $n-1$ unit $(s,t)$-paths.
Every source arc and every sink arc belongs to exactly one of
these paths. Thus $Q_v$ and $g_v$ are well-defined for all
$v\in V\setminus\{r\}$, and the assignment $v\mapsto g_v$ is
a bijection, proving part~(\ref{prop:selected-edges:bijection}).

Let $v\in V\setminus\{r\}$. If $Q_v$ avoids $h$, the construction
of $\mathcal N(T)$ ensures that its graph edge $e$ is incident
with $v$ and that $g_v\in P_T(e)$. Choosing $w_v$ as the other
end-vertex of $e$ therefore gives $g_v\in P_T(v,w_v)$. If $Q_v$
passes through $h$, exactly one end-vertex of $g_v$ lies in the
component of $T-g_v$ not containing $v$. This defines $w_v$ and
again gives $g_v\in P_T(v,w_v)$. Hence every selected copy is
well-defined, and part~(\ref{prop:selected-edges:path}) holds.
\end{proof}

\paragraph{Two-disjoint spanning trees.}
Next, we show that the multiset $F\uplus E(T)$ contains two disjoint spanning trees. We emphasize that the proof of this does not require $f$ to be a max flow in $\mathcal{N}(T)$. 
\begin{lemma}\label{lem:two-tree-decomposition}
Let $T$ be a tree on $V$ rooted at $r$. For each $v\in V\setminus\{r\}$, let
$e_v=\{v,w_v\}$ be an edge copy, and let $F$ be the multiset of these
copies. Suppose that there is a bijection $v\mapsto g_v$ from
$V\setminus\{r\}$ to $E(T)$ such that $g_v\in P_T(v,w_v)$
for all $v\in V\setminus\{r\}$.
Then, the graph $(V, E(T)\uplus F)$ can be partitioned into two edge-disjoint spanning trees.
\end{lemma}
\begin{proof}
Orient every edge of $T$ away from $r$. For all
$v\in V\setminus\{r\}$, orient the copy $e_v=vw_v$ from $w_v$ to $v$. Let $H$ be the resulting directed multigraph.
We show that every nonempty set
$S\subseteq V\setminus\{r\}$ has at least two arcs entering it in $H$.
In the orientation of $T$, each component of $T[S]$ has exactly one
entering arc. If $T[S]$ has at least two components, the claim follows
already from the tree arcs. Otherwise, $T[S]$ is connected and exactly
one tree arc enters $S$. Suppose no arc from $F$ enters $S$. Since
the copy $e_v$ is directed into $v$, this would imply $w_v\in S$ for
every $v\in S$. Connectivity of $T[S]$ then gives, for all $v\in S$,
\[
 g_v\in P_T(v,w_v)\subseteq E(T[S]).
\]
The bijection would assign $|S|$ distinct tree edges to a set of
only $|S|-1$ edges, a contradiction. Thus a second arc enters $S$.
Edmonds' arborescence packing theorem~\cite{Edm73} now guarantees
two arc-disjoint spanning arborescences rooted at $r$ and directed
away from $r$. Together they contain $2n-2$ arcs, which is all of $H$.
Forgetting directions therefore partitions the edge copies of
$E(T)\uplus F$ into two spanning trees, as required.
\end{proof}

We now state the algorithm and prove its correctness. See Algorithm \ref{alg:excess-halving} for the pseudocode.  

\begin{algorithm}[H]
\caption{\textsc{HalveExcess}$(G,T,r)$}
\label{alg:excess-halving}
\textbf{Input:} A graph $G=(V,E,q)$, a tree $T$ on $V$ with
$|V|\ge2$, and a root $r\in V$.\\
\textbf{Output:} A tree $T'$ satisfying~\eqref{eq:pps-update-halving}
for every fixed optimum partition $\Pi^*$.
\begin{enumerate}
 \item Compute a rational maximum $(s,t)$-flow $f$ in $\mathcal N(T)$.
 \item Construct $\augment(T)$ and $f^{\mathrm{aug}}$ as in
 Definition~\ref{def:augment}.
 \item Apply Lemma~\ref{lem:flow-rounding} to $f^{\mathrm{aug}}$
 to obtain a randomized integral flow $f^{\mathrm{aug}}_{\mathrm{int}}$ in $\augment(T)$.
 \item Decompose $f^{\mathrm{aug}}_{\mathrm{int}}$ into unit paths
 and construct $F$ as in Definition~\ref{def:selected-edges}.
 \item Partition the multiset $E(T)\uplus F$ into two spanning trees $T_1,T_2$, (which exists via Proposition \ref{prop:selected-edges} and Lemma \ref{lem:two-tree-decomposition}). 
 \item Choose $j\in\{1,2\}$ uniformly at random and return $T_j$.
\end{enumerate}
\end{algorithm}

\paragraph{Correctness.} We now prove the excess-halving guarantee of Algorithm \ref{alg:excess-halving}. This argument will use
the fact that $f$ is a maximum flow in $\mathcal N(T)$.
We recall that $D(\Pi)=|\Pi|-1-q(\delta_G(\Pi))$
for every vertex partition $\Pi$, and $D^*=D(\Pi^*)$.

\begin{lemma}\label{lem:halving-correctness}
Algorithm~\ref{alg:excess-halving} returns a tree $T'$ on $V$. Moreover, for every fixed optimum partition $\Pi^*$ of the Lagrangian problem on $G$, we have that 
$\mathbb E[z_{\Pi^*}(T')]\le\tfrac12 z_{\Pi^*}(T)$.
\end{lemma}
\begin{proof}
Proposition~\ref{prop:selected-edges} ensures that the multiset $F$
as constructed in Definition~\ref{def:selected-edges} is well-defined.
Part~(\ref{prop:selected-edges:bijection}) gives the bijection
$v\mapsto g_v$, and part~(\ref{prop:selected-edges:path}) gives
the path condition~\eqref{eq:selected-path-property}.
Lemma~\ref{lem:two-tree-decomposition} therefore guarantees that
the algorithm returns a tree. We next bound its expected excess.

Fix a maximum flow $f$ chosen by the algorithm. Let $\augflow$ be obtained from $f$ as in Definition \ref{def:augment} and let $\intaugflow$ be obtained from $\augflow$ by applying Lemma \ref{lem:flow-rounding}. 
For every $e\in E$, let $X_e$
be the number of copies of $e$ selected in $F$ by paths avoiding $h$.
Each such path traverses $y_e^-\to y_e^+$ exactly once. Thus, for
all $e\in E$, Proposition~\ref{prop:rounded-flow}(\ref{prop:rounded-flow:edge}) gives
\begin{align*}
 X_e&=\intaugflow(y_e^-,y_e^+),\\
 \mathbb E[X_e]&\le q_e.
\end{align*}
Let $K$ be the number of dummy edges selected. Each path through
$h$ uses exactly one arc $a_v\to h$. Hence
Proposition~\ref{prop:rounded-flow}(\ref{prop:rounded-flow:dummy}) gives
\begin{align*}
 K&=\sum_{v\in V\setminus\{r\}}\intaugflow(a_v,h),\\
 \mathbb E[K]&=\alpha_T.
\end{align*}
By Definition~\ref{def:selected-edges}, an edge-copy selected by a path avoiding $h$ crosses $\Pi^*$ only if its
graph edge belongs to $\delta_G(\Pi^*)$. 
Hence, 
\begin{align}
 \mathbb E[|\delta_F(\Pi^*)|]
 &\le \sum_{e\in\delta_G(\Pi^*)}\mathbb E[X_e]+\mathbb E[K]
 \notag\\
 &\le q(\delta_G(\Pi^*))+\alpha_T
 \notag\\
 &\le q(\delta_G(\Pi^*))+D^*
 =|\Pi^*|-1.\label{eq:selected-crossing-bound}
\end{align}
The last inequality above is by Lemma~\ref{lem:pps-cut-deficit}. 

For every realized $F$, the two trees partition $E(T)\uplus F$.
Consequently, their total number of crossings of $\Pi^*$ equals
$|\delta_T(\Pi^*)|+|\delta_F(\Pi^*)|$. Uniformly choosing a tree
and using~\eqref{eq:selected-crossing-bound} gives
\begin{align*}
 \mathbb E[z_{\Pi^*}(T')]
 &=\frac12\bigl(|\delta_T(\Pi^*)|
             +\mathbb E[|\delta_F(\Pi^*)|]\bigr)-(|\Pi^*|-1)\\
 &\le\frac12\bigl(|\delta_T(\Pi^*)|-(|\Pi^*|-1)\bigr)
 =\frac12 z_{\Pi^*}(T).
\end{align*}
\end{proof}

\subsection{Run-time}\label{sec:halving-runtime}
We implement Algorithm~\ref{alg:excess-halving} using the compact
network $\compact(T)$ from Section~\ref{sec:flow}.
The following lemma gives the required implementation and its run-time.

\begin{lemma}\label{lem:halving-runtime}
Suppose $n\ge2$. Algorithm~\ref{alg:excess-halving}
can be implemented with the following running-time guarantees:
\begin{enumerate}
\item The algorithm uses one maximum-flow computation on
a network with $O(m+n)$ vertices and $O(n+m\log^2 n)$ arcs,
where every arc capacity belongs to
$\{1,n\}\cup\{q_e:e\in E\}$, together with
\[
    O\!\left((n+m\log^2 n)\log(m+n)\right)
\]
additional expected arithmetic operations.

\item Let $S\ge m+n$ and let $C>0$ be fixed.
Suppose that $q_e=a_e/B$ for every $e\in E$, where $B$
and all $a_e$ are positive integers satisfying
$B\le S^C$ and $a_e\le S^C$ for every $e\in E$.
If the maximum-flow computation returns a flow whose arc
values are integer multiples of $1/B$, then the additional
expected running time is
\[
    O\!\left((m+n)\log^{O(1)}S\right),
\]
where the exponent depends only on $C$.
\end{enumerate}
\end{lemma}

\begin{proof}
We first describe the implementation. Construct $\compact(T)$
using Lemma~\ref{lem:pps-compact-network}, implementing every
infinite-capacity arc with capacity $n$, and compute a rational
maximum $(s,t)$-flow $f_{\mathrm c}$ in this network.
We define the augmented compact network and its flow explicitly.
Let $\mathcal N_{\mathrm c}^{\mathrm{aug}}(T)$ be obtained from
$\compact(T)$ by adding a dummy node $h$, unit-capacity arcs
$a_v\to h$ for all $v\in V\setminus\{r\}$, and unit-capacity
arcs $h\to z_g$ for all $g\in E(T)$.
For every $v\in V\setminus\{r\}$, define
\begin{align*}
 f_{\mathrm c}^{\mathrm{aug}}(s,a_v)&=1,\\
 f_{\mathrm c}^{\mathrm{aug}}(a_v,h)&=1-f_{\mathrm c}(s,a_v).
\end{align*}
For every $g\in E(T)$, define
\begin{align*}
 f_{\mathrm c}^{\mathrm{aug}}(z_g,t)&=1,\\
 f_{\mathrm c}^{\mathrm{aug}}(h,z_g)&=1-f_{\mathrm c}(z_g,t).
\end{align*}
On every other arc, retain the value of $f_{\mathrm c}$.
Apply Lemma~\ref{lem:flow-rounding} to
$f_{\mathrm c}^{\mathrm{aug}}$ to obtain an integer-valued
flow $f^{\mathrm{aug}}_{\mathrm{c,int}}$ on the same arcs
satisfying the lemma's rounding guarantees, and decompose
it into $n-1$ unit $(s,t)$-paths.

Select edge copies as in Definition~\ref{def:selected-edges},
with the following interpretation of paths in the compact
network. For every $v\in V\setminus\{r\}$, let $Q_v$ be the
path starting with $s\to a_v$, and let $z_{g_v}$ be its last
internal node. If $Q_v$ avoids $h$, read its arc
$y_e^-\to y_e^+$, let $w_v$ be the other end-vertex of the
graph edge $e$ incident with $v$, and select a copy
$e_v=\{v,w_v\}$ of $e$. If $Q_v$ passes through $h$, let
$w_v$ be the end-vertex of $g_v$ in the component of
$T-g_v$ not containing $v$, and select a dummy copy
$e_v=\{v,w_v\}$. Let $F$ be the multiset of the selected
copies, retaining the assignment $v\mapsto(e_v,g_v)$.
Find two edge-disjoint spanning trees in
$(V,E(T)\uplus F)$, choose one uniformly at random, and
return it.

We next prove correctness. The source cut has capacity
$n-1$, whereas any cut crossing an arc whose infinite
capacity was replaced by $n$ has capacity at least $n$.
Thus this replacement preserves the minimum cuts and
their capacities. By the flow-preservation property of
Lemma~\ref{lem:pps-compact-network}, the value of
$f_{\mathrm c}$ is $f_T$.
The same conservation calculation as in the proof of
Proposition~\ref{prop:augmented-flow} shows that
$f_{\mathrm c}^{\mathrm{aug}}$ is feasible in
$\mathcal N_{\mathrm c}^{\mathrm{aug}}(T)$ and has value
$n-1$. Moreover,
\[
 f_{\mathrm c}^{\mathrm{aug}}(y_e^-,y_e^+)
 =f_{\mathrm c}(y_e^-,y_e^+)\le q_e
 \qquad\text{for every }e\in E,
\]
and
\[
 \sum_{v\in V\setminus\{r\}}
 f_{\mathrm c}^{\mathrm{aug}}(a_v,h)
 =n-1-f_T=\alpha_T.
\]
Consequently, the proof of
Proposition~\ref{prop:rounded-flow} applies to this
augmented compact flow as well: the rounded flow has
value $n-1$, every source and sink arc carries one unit,
each graph edge arc has expected flow at most $q_e$,
and the expected total flow through $h$ is $\alpha_T$.
The augmented compact network is acyclic, so the rounded
flow decomposes into $n-1$ unit $(s,t)$-paths, and
$v\mapsto g_v$ is a bijection.
For a path avoiding $h$, the segment from $y_e^+$ to
$z_{g_v}$ runs through the auxiliary forest. By the
reachability property of
Lemma~\ref{lem:pps-compact-network},
\[
    g_v\in P_T(v,w_v).
\]
For a path through $h$, the same condition follows from
the choice of $w_v$. These are the two properties in
Proposition~\ref{prop:selected-edges}, so
Lemma~\ref{lem:two-tree-decomposition} guarantees the
required two edge-disjoint spanning trees.

For every $e\in E$, let $X_e$ be the number of copies of
$e$ selected by paths avoiding $h$, and let $K$ be the
number of dummy copies. Each selected graph edge copy
corresponds to one unit of flow through its graph edge
arc, and each dummy copy corresponds to one unit of
flow through $h$. The rounding guarantees therefore give
\begin{align*}
    \mathbb E[X_e]&\le q_e \text{ for every }e\in E,\text{ and}\\
    \mathbb E[K]&=\alpha_T.
\end{align*}
These are precisely the estimates used in
\eqref{eq:selected-crossing-bound}.
Since $f_{\mathrm c}$ has maximum-flow value $f_T$,
Lemma~\ref{lem:pps-cut-deficit} gives $\alpha_T\le D^*$.
The averaging argument in the proof of
Lemma~\ref{lem:halving-correctness} therefore proves the
same excess-halving guarantee for the compact
implementation.

Finally, we analyze the running time.
By Lemma~\ref{lem:pps-compact-network}, constructing
$\compact(T)$ takes $O(n+m\log^2 n)$ arithmetic operations.
Its retained arcs have capacities $1$, $q_e$, or $n$,
and every auxiliary arc has capacity $n$.
Thus the single maximum-flow computation uses a network
whose arc capacities belong to
\[
    \{1,n\}\cup\{q_e:e\in E\}.
\]
Adding the dummy node and arcs and completing the flow
takes $O(n)$ additional arithmetic operations.
Let $N$ and $M$ be the numbers of vertices and arcs in
the augmented compact network. Then
$N=O(m+n)$ and
$M=O(n+m\log^2 n)$.
To implement Lemma~\ref{lem:flow-rounding}, add a return
arc $t\to s$ carrying $n-1$ units and apply the unbiased
circulation-rounding algorithm of Kang and
Payor~\cite[Sections~3.3 and~4]{KP15}, using the initial
floor and ceiling of each arc flow as its rounding bounds.
It takes $O(M\log N)$ expected arithmetic operations.
The return arc retains its integer value by
\eqref{eq:rounding-bounds}; removing it gives
$f^{\mathrm{aug}}_{\mathrm{c,int}}$.

Every $(s,t)$-path avoiding $h$ has $O(\log n)$ arcs
by Lemma~\ref{lem:pps-compact-network}, and every path
through $h$ has constant length.
For every $v\in V\setminus\{r\}$, start with $s\to a_v$,
follow arcs with positive remaining integral flow until
reaching $t$, and subtract one unit along the resulting
path. Conservation and acyclicity ensure that this
procedure reaches $t$.
Maintain adjacency-list pointers that skip arcs once
their remaining flow becomes zero. Each arc is skipped
at most once, and the $n-1$ extracted paths have total
length $O(n\log n)$, so the decomposition takes
$O(M+n\log n)$ arithmetic operations.
For a path avoiding $h$, read $e_v$ and $g_v$ directly
from the path. For a dummy path, find $w_v$ in constant
time after $O(n)$ preprocessing of $T$: root $T$ at $r$
and record the depth-first entry and exit times of each
vertex. If $g_v$ joins a parent $x$ to a child $y$,
choose $w_v=x$ when $v$ belongs to the subtree rooted
at $y$, and choose $w_v=y$ otherwise.
Hence constructing $F$ and its assignments requires
no additional asymptotic work.

Construct the orientation $H$ from the proof of
Lemma~\ref{lem:two-tree-decomposition} in $O(n)$ time.
It has $n$ vertices and $2n-2$ arcs.
Tarjan's algorithm~\cite{Tar76} finds two arc-disjoint
spanning arborescences rooted at $r$ and directed away
from $r$ in $O(n\alpha(n))$ time, where $\alpha$ is an
inverse-Ackermann function.
Forgetting their orientations gives the two required
spanning trees. Selecting and returning one takes
$O(n)$ time.
Consequently, the total work beyond the one maximum-flow
computation is
\[
\begin{aligned}
 &O(n+m\log^2 n)+O(M\log N)\\
 &\qquad{}+O(M+n\log n)+O(n\alpha(n))\\
 &\quad=O\!\left((n+m\log^2 n)\log(m+n)\right)
\end{aligned}
\]
expected arithmetic operations, proving the first guarantee.

For the second guarantee, fix $S\ge m+n$ and a constant
$C>0$. Suppose that $q_e=a_e/B$ for every $e\in E$,
where $B$ and all $a_e$ are positive integers at most
$S^C$, and that the returned maximum flow has arc values
in $(1/B)\mathbb Z$.
Represent capacities and flow values using their integer
numerators over the common denominator $B$.
The capacity numerators belong to
\[
    \{B,nB\}\cup\{a_e:e\in E\},
\]
so they are at most $S^{C+1}$.
The completion formulas preserve membership in
$(1/B)\mathbb Z$, as does adding the return arc.
Since the augmented compact network is acyclic and its
flow has value $n-1$, every arc flow before adding the
return arc lies between $0$ and $n-1$.
The return arc also carries $n-1$ units.
Thus every initial flow numerator is at most
$(n-1)B\le S^{C+1}$.

During cycle-cancelling rounding, each update amount
is a minimum of distances from current arc flows to
their fixed integer rounding bounds. Inductively, all
update amounts and all updated flow values remain in
$(1/B)\mathbb Z$.
The flows remain between their initial floors and ceilings,
and therefore remain between $0$ and $n-1$.
Their numerators over $B$ consequently have $O(\log S)$
bits, with the implied constant depending only on $C$.
The availabilities have numerators at most $B$.
The dynamic-tree implementation maintains these quantities
using comparisons, minima, and additions or subtractions
of update amounts. Its stored numerical quantities can
be represented with numerators of magnitude polynomial
in $N$, $M$, $n$, and $B$.
Since $N$ and $M$ are polynomially bounded in $S$,
these quantities also have $O(\log S)$ bits.

It remains to account for the random choices in rounding.
If the available update amounts in the two directions
of a fractional cycle are $d^+/B$ and $d^-/B$, then the
forward direction is chosen with probability
\[
    \frac{d^-}{d^++d^-}.
\]
Here $d^+$ and $d^-$ are positive integers at most $B$.
Sample an integer uniformly from
$\{1,\ldots,d^++d^-\}$ and choose the forward direction
exactly when the sampled integer is at most $d^-$.
Rejection sampling with independent unbiased bits
implements this choice exactly. Each trial uses
$O(\log S)$ bits and succeeds with probability at least
$1/2$, so each choice takes $\log^{O(1)}S$ expected time.
The final uniform choice between the two spanning trees
requires one unbiased bit.

All remaining steps manipulate flow numerators of
$O(\log S)$ bits, graph indices, and counters whose
magnitudes are polynomial in $S$.
Thus each arithmetic operation and each random choice
has $\log^{O(1)}S$ expected cost.
Combining this with the first guarantee, the total
expected running time excluding the maximum-flow
computation is
\[
\begin{aligned}
 &O\!\left(
   (n+m\log^2 n)\log(m+n)\log^{O(1)}S
 \right)\\
 &\qquad=
 O\!\left((m+n)\log^{O(1)}S\right),
\end{aligned}
\]
where the exponent depends only on $C$.
\end{proof}

We now complete the proof of Lemma \ref{lem:halving}. 
\begin{proof}[Proof of Lemma~\ref{lem:halving}]
We first describe the algorithm. For $n=1$, return the tree
with no edges. For $n\ge2$, run
Algorithm~\ref{alg:excess-halving} using the compact-network
implementation of Lemma~\ref{lem:halving-runtime}.

Next, we prove correctness. For $n=1$, the excess is zero.
For $n\ge2$, Lemma~\ref{lem:halving-runtime} provides an
implementation of Algorithm~\ref{alg:excess-halving}, so
Lemma~\ref{lem:halving-correctness} implies that the returned
tree satisfies~\eqref{eq:pps-update-halving} for every fixed
optimum partition $\Pi^*$.

Finally, we prove the running-time guarantees.
The case $n=1$ takes constant time and requires no
maximum-flow computation. Suppose that $n\ge2$.
By the first guarantee of Lemma~\ref{lem:halving-runtime},
the implementation uses one maximum-flow computation on
a network with $O(m+n)$ vertices and $O(n+m\log^2 n)$ arcs,
where every arc capacity belongs to
\[
    \{1,n\}\cup\{q_e:e\in E\},
\]
together with
\[
    O\!\left((n+m\log^2 n)\log(m+n)\right)
\]
additional expected arithmetic operations.
For the second guarantee, fix $S\ge m+n$ and a constant
$C>0$. Suppose that $q_e=a_e/B$ for every $e\in E$,
where $B$ and all $a_e$ are positive integers at most
$S^C$, and that the maximum-flow computation returns
a flow whose arc values are integer multiples of $1/B$.
The second guarantee of Lemma~\ref{lem:halving-runtime}
then applies with the same $S$ and $C$, giving
\[
    O\!\left((m+n)\log^{O(1)}S\right)
\]
expected time excluding the maximum-flow computation,
where the exponent depends only on $C$.
\end{proof}

\section{Algorithm for the Lagrangian Problem}
\label{sec:lagrangian}

In this section, we prove that the Lagrangian problem
\eqref{eq:lagrangian} can be solved in almost linear-time
and use it to prove Theorem \ref{thm:oracle}. 
Combining Lemma~\ref{thm:pps-find-respecting-tree} with
Lemma~\ref{lem:pps-recover-optimum} gives the following result.

\begin{lemma}\label{lem:solving-the-lagrangian}
Given an undirected graph
$G=(V,E,c:E\rightarrow\Z_+)$ with $n\ge1$ vertices,
$m$ edges, and positive integer capacities, a rational
$\lambda\ge0$, and $\eta\in(0,1)$, there exists a randomized
algorithm that returns a partition minimizing
$\Phi_{G,\lambda}$, together with its objective value,
with probability at least $1-\eta$.
The algorithm has the following running-time guarantees:
\begin{enumerate}
\item If $n=1$ or $\lambda=0$, the algorithm returns
$\{V\}$ with objective value zero deterministically
in $O(n)$ time and requires no maximum-flow computation.

\item If $n\ge2$ and $\lambda>0$, the algorithm uses
$O(\log(n/\eta))$ maximum-flow computations, each on
a network with $O(m+n)$ vertices and
$O((m+n)\log^2 n)$ arcs, where every arc capacity belongs
to $\{1,n\}\cup\{c_e/\lambda:e\in E\}$, together with
\[
    O\!\left(
        (m+n)\log^2 n\log(m+n)\log(n/\eta)
    \right)
\]
additional expected arithmetic operations.

\item Suppose that $n\ge2$ and $\lambda=p/d>0$, where
$p,d$ are positive integers. Let $S\ge m+n$ and let
$C>0$ be fixed. Suppose that
$c_e\le S^C$  for every $e\in E$, 
and that $p$, $d$, and $1/\eta$ are at most $S^C$.
If every maximum-flow computation returns a flow whose
arc values are integer multiples of $1/p$, then the
additional expected running time is
\[
    O\!\left((m+n)\log^{O(1)}S\right),
\]
where the exponent depends only on $C$.
\end{enumerate}
\end{lemma}

\begin{proof}
We first describe the algorithm. If $n=1$ or $\lambda=0$,
return the partition $\{V\}$ with objective value zero.
Otherwise, construct $H=(V,E,q)$ by setting
$q_e=c_e/\lambda$ for every $e\in E$.
Apply Lemma~\ref{thm:pps-find-respecting-tree} to $H$
with failure parameter $\eta$ to obtain a tree $T$.
Run the recovery algorithm of
Lemma~\ref{lem:pps-recover-optimum} on $H$ and $T$
to obtain a partition $\Pi$.
Compute $\Phi_{G,\lambda}(\Pi)$ by scanning the graph
edges, and return $\Pi$ and this objective value.

Next, we prove correctness. If $n=1$, there is only one
partition, and its objective value is zero.
If $\lambda=0$, every partition has nonnegative objective
value, and $\{V\}$ has value zero.
Now suppose that $n\ge2$ and $\lambda>0$.
For every partition $\Pi$,
\[
\begin{aligned}
    \Phi_{H,1}(\Pi)
    &=q(\delta_G(\Pi))-(|\Pi|-1)
    =\frac{\Phi_{G,\lambda}(\Pi)}{\lambda}.
\end{aligned}
\]
Since $\lambda>0$, the optimum partitions for $(G,\lambda)$
and $(H,1)$ coincide.
Fix an optimum partition $\Pi^*$ for $(H,1)$ independently
of the algorithm's random choices; this partition is used
only for the analysis.
By Lemma~\ref{thm:pps-find-respecting-tree}, the tree $T$
respects $\Pi^*$ with probability at least $1-\eta$.
On this event, Lemma~\ref{lem:pps-recover-optimum}
guarantees that the recovered partition $\Pi$ is optimal
for $(H,1)$ and hence for $(G,\lambda)$.
The final scan computes its objective value exactly,
establishing the claimed success probability.

Finally, we prove the running-time guarantees.
If $n=1$ or $\lambda=0$, explicitly outputting the
partition $\{V\}$ takes $O(n)$ time, and no maximum-flow
computation is needed.
Suppose that $n\ge2$ and $\lambda>0$.
Normalizing the capacities takes $O(m)$ arithmetic
operations.
By Lemma~\ref{thm:pps-find-respecting-tree}, finding $T$
uses $O(\log(n/\eta))$ maximum-flow computations.
The recovery algorithm of
Lemma~\ref{lem:pps-recover-optimum} uses one additional
maximum-flow computation.
Its construction and running-time analysis apply to
every tree $T$; the respecting property is needed only
to guarantee optimality of the recovered partition.
Each maximum-flow network has $O(m+n)$ vertices and
$O((m+n)\log^2 n)$ arcs, with every arc capacity belonging
to
\[
    \{1,n\}\cup\{q_e:e\in E\}
    =
    \{1,n\}\cup\{c_e/\lambda:e\in E\}.
\]
Assigning a part label to each vertex and scanning the
graph edges to evaluate $\Phi_{G,\lambda}(\Pi)$ takes
$O(m+n)$ additional arithmetic operations.
Combining these costs with the additional-operation
bounds in the two lemmas gives
\[
    O\!\left(
        (m+n)\log^2 n\log(m+n)\log(n/\eta)
    \right)
\]
additional expected arithmetic operations, proving
the second guarantee.

For the third guarantee, fix $S\ge m+n$ and a constant
$C>0$. Suppose that $\lambda=p/d>0$, where all capacities
$c_e$ and the positive integers $p,d$ are at most $S^C$,
that $1/\eta\le S^C$, and that every maximum-flow
computation returns arc values in $(1/p)\mathbb Z$.
For every $e\in E$,
\[
    q_e=\frac{dc_e}{p}.
\]
Thus the normalized capacities have common denominator
$B:=p$, with $B\le S^C$ and $dc_e\le S^{2C}$ for every $e\in E$.
Consequently, the second guarantees of
Lemmas~\ref{thm:pps-find-respecting-tree}
and~\ref{lem:pps-recover-optimum} apply with the same
size parameter $S$ and exponent $2C$.
They bound the total expected running time of tree finding
and partition recovery, excluding maximum-flow computations,
by
\[
    O\!\left((m+n)\log^{O(1)}S\right),
\]
where the exponent depends only on $C$.

Normalizing the capacities also obeys this bound:
the integers $p$, $d$, and $dc_e$ have $O(\log S)$ bits.
For objective evaluation, use
\[
    \Phi_{G,\lambda}(\Pi)
    =
    \frac{d\,c(\delta_G(\Pi))-p(|\Pi|-1)}{d}.
\]
Since $m,n\le S$, every integer involved in this
computation has $O(\log S)$ bits, with the implied
constant depending only on $C$.
Thus normalization and objective evaluation together take
$O((m+n)\log^{O(1)}S)$ time.
The total expected running time excluding maximum-flow
computations is therefore
\[
    O\!\left((m+n)\log^{O(1)}S\right),
\]
as claimed.
\end{proof}

In order to prove Theorem \ref{thm:oracle}, it remains to show that the unique finest optimum to the
Lagrangian can be obtained in almost-linear time. 
Let $G=(V,E,c: E\rightarrow \Z_+)$ be the input graph with positive integer
capacities. 
We recall that at every $\lambda$, there exists a unique finest optimal partition to the Lagrangian problem \cite{Nar91, Nar97}. 
We show that a small perturbation of $\lambda$
makes the optimum to the Lagrangian unique and recovers the finest partition.

\begin{lemma}
\label{lem:pps-perturbation}
Let $\lambda=p/d\ge0$, where $p\ge0$ and $d\ge1$ are
integers, and define
\[
 \lambda^+
 :=\lambda+\frac{1}{2dn}
 =\frac{2np+1}{2nd}.
\]
The Lagrangian problem at $\lambda^+$ has a unique
optimum partition and it is the finest optimum partition for the Lagrangian problem at $\lambda$.
\end{lemma}

\begin{proof}
Let $\Pi$ and $\mathcal{Q}$ be two partitions of $V$ such that
$\Phi_{G,\lambda}(\Pi)>\Phi_{G,\lambda}(\mathcal{Q})$.
Since the capacities are integral and $\lambda=p/d$, every
objective value at $\lambda$ is an integer multiple of $1/d$.
Consequently,
\[
 \Phi_{G,\lambda}(\Pi)-\Phi_{G,\lambda}(\mathcal{Q})
 \ge \frac1d.
\]
By the definition of $\lambda^+$, for every partition $\Pi$,
\[
 \Phi_{G,\lambda^+}(\Pi)
 =
 \Phi_{G,\lambda}(\Pi)-\frac{|\Pi|-1}{2dn}.
\]
Since the numbers of parts of two partitions differ by at most
$n-1$, we obtain
\begin{align*}
 \Phi_{G,\lambda^+}(\Pi)-\Phi_{G,\lambda^+}(\mathcal{Q})
 &=
 \Phi_{G,\lambda}(\Pi)-\Phi_{G,\lambda}(\mathcal{Q})
 -\frac{|\Pi|-|\mathcal{Q}|}{2dn}\\
 &\ge \frac1d-\frac{n-1}{2dn}\\
 &>0.
\end{align*}
Thus every partition that is not optimal at $\lambda$ remains
strictly worse at $\lambda^+$ than every partition that is optimal
at $\lambda$. In particular, every optimum partition at
$\lambda^+$ is also an optimum partition at $\lambda$.

Now let $\Pi$ and $\mathcal{Q}$ be optimum partitions at $\lambda$.
Their objective values at $\lambda$ are equal, so
\[
 \Phi_{G,\lambda^+}(\Pi)-\Phi_{G,\lambda^+}(\mathcal{Q})
 =
 -\frac{|\Pi|-|\mathcal{Q}|}{2dn}.
\]
Hence, among optimum partitions at $\lambda$, the objective at
$\lambda^+$ is minimized exactly by those with the greatest
number of parts. It is known that there is a unique optimum
partition with the greatest number of parts~\cite{Nar91,Nar97}. Together with the preceding
paragraph, this shows that it is the unique optimum partition
at $\lambda^+$.
\end{proof}

Lemmas~\ref{lem:solving-the-lagrangian}
and~\ref{lem:pps-perturbation} now complete the proof
of Theorem~\ref{thm:oracle}.

\begin{proof}[Proof of Theorem~\ref{thm:oracle}]
We first describe the algorithm. If $n=1$, return the partition
$\{V\}$ with objective value zero. Henceforth, assume that $n\ge2$.
Write $\lambda=p/d$, where $p\ge0$ and $d\ge1$ are integers, and set
\[
 \lambda^+:=\lambda+\frac{1}{2dn}
 =\frac{2np+1}{2nd}.
\]
Apply the algorithm of Lemma~5.1 to $G$ with parameter
$\lambda^+$ and error probability $\eta$, obtaining a partition
$\Pi$. Compute $\Phi_{G,\lambda}(\Pi)$ by scanning the graph edges,
and return $\Pi$ together with this value.

Next, we prove correctness. By Lemma~5.1, with probability at least
$1-\eta$, the returned partition $\Pi$ is optimal at $\lambda^+$.
By Lemma~5.2, the optimum at $\lambda^+$ is unique and is the finest
optimum at $\lambda$. Thus, with probability at least $1-\eta$,
the algorithm returns the required partition and its objective
value at the original parameter $\lambda$.

Finally, we analyze the run-time. Since $\lambda^+>0$, Lemma~5.1
gives $O(\log(n/\eta))$ maximum-flow computations, each on a network
with $O(m+n)$ vertices and $O((m+n)\log^2 n)$ arcs, together with
\[
 O\!\left((m+n)\log^2 n\log(m+n)\log(n/\eta)\right)
\]
additional expected arithmetic operations.
Forming $\lambda^+$ requires a constant number of arithmetic
operations, and evaluating $\Phi_{G,\lambda}(\Pi)$ requires
$O(m+n)$ additional operations. These costs are absorbed by the
stated bound.

Now suppose that the capacities $c_e$ and the integers $p,d$
are polynomially bounded in $m+n$, and that $\eta=n^{-a}$
for a fixed $a>0$. Let $B:=2np+1$.
By the capacity guarantee in Lemma~5.1, every maximum-flow
network used at parameter $\lambda^+$ has arc capacities in
\[
 \{1,n\}\cup
 \left\{\frac{c_e}{\lambda^+}:e\in E\right\}
 =
 \{1,n\}\cup
 \left\{\frac{2ndc_e}{B}:e\in E\right\}.
\]
Multiplying all arc capacities by $B$ produces integer capacities in
\[
 \{B,nB\}\cup\{2ndc_e:e\in E\},
\]
all polynomially bounded in $m+n$.
We compute an integral maximum flow in the scaled network
using~\cite{BCKLPPSS23} and divide its arc values by $B$.
This yields a maximum flow in the original network whose arc
values are integer multiples of $1/B$.
Scaling, solving, and rescaling each instance takes
$(m+n)^{1+o(1)}$ time.
Using the representation $\lambda^+=B/(2nd)$, the
additional-running-time guarantee in Lemma~5.1 applies:
both $B$ and $2nd$ are polynomially bounded, and every returned
maximum flow has arc values that are integer multiples of $1/B$.
Consequently, all work outside the maximum-flow computations
takes $\widetilde O(m+n)$ expected time.
Evaluating the returned partition at the original parameter
also takes $\widetilde O(m+n)$ time.
Since there are $O(\log n)$ maximum-flow computations, the total
expected running time is $(m+n)^{1+o(1)}$.
\end{proof}

\section{Fast canonical cut hierarchy}
\label{sec:cch}

In this section, we prove Theorem~\ref{thm:cch}
using Theorem~\ref{thm:oracle}.
We recall the definition of the canonical cut hierarchy.
For a connected graph $G$, its root represents $V(G)$.
An internal node representing $S$ is labelled by $\beta(G[S])$,
and its children represent the parts of the finest
Tutte--Nash--Williams partition of $G[S]$.
Singletons are leaves.
For a disconnected graph, the hierarchies of its connected
components are attached to a root labelled zero.
We store child pointers and labels so the set represented by a
node is specified implicitly by its descendant leaves.

Throughout this section, capacities are positive integers.
For a graph $H$ and $\lambda\ge 0$, let $\Pi_H(\lambda)$
denote the unique finest partition minimizing
$\Phi_{H,\lambda}$, and let
$\mathcal S_H=\bigl\{\{v\}:v\in V(H)\bigr\}$
denote its singleton partition.
Theorem~\ref{thm:oracle} provides the algorithm for computing
$\Pi_H(\lambda)$.

In Section \ref{sec:lagrangian-optima-properties}, we recall the classical properties of Lagrangian optima that will be central to our algorithm. 
In Section \ref{sec:cch-algo}, we describe the algorithm. We prove correctness in Section \ref{sec:cch-correctness} and bound runtime in Section \ref{sec:cch-runtime}. 

\subsection{Properties of the Lagrangian optima}\label{sec:lagrangian-optima-properties}
Let $H=(V, E, c: E\rightarrow \Z_+)$ be an undirected graph. 
For a partition $\mathcal P$ of $V$, let $H/\mathcal P$
be obtained by contracting each part to a vertex and deleting
loops. We retain parallel edges with their original capacities
and identify each contracted vertex with the part it represents.
For a partition $\mathcal Q$ of $V(H/\mathcal P)$, we define its
\emph{lift} to $V(H)$ by
\[
 \operatorname{lift}_{\mathcal P}(\mathcal Q)
 :=
 \left\{
   \bigcup_{B\in X} B : X\in\mathcal Q
 \right\}.
\]

It is known that as the parameter $\lambda$ increases, the finest optimum only splits its
parts. We summarise this in the lemma below. This lemma is standard---e.g., see \cite{Nar97, RS08}. We include a proof for completeness. 

\begin{lemma}
\label{lem:cch-restriction-contraction} 
Let $H=(V, E, c: E\rightarrow \Z_+)$ be an undirected graph. The following properties hold.
\begin{enumerate}
 \item For every $0\le x\le y$, the partition $\Pi_H(y)$
 refines $\Pi_H(x)$.
 \item For every $\lambda\ge0$ and every
 $B\in\Pi_H(\lambda)$, the induced graph $H[B]$ is connected.
 \item For a fixed $t\ge0$, let $\mathcal P=\Pi_H(t)$. Then, 
 \[
 \begin{aligned}
  \Pi_H(\lambda)
  &=
  \begin{cases}
  \operatorname{lift}_{\mathcal P}
    \bigl(\Pi_{H/\mathcal P}(\lambda)\bigr)
  &\ \forall\ \lambda\in[0,t],\\
  \bigcup_{B\in\mathcal P}\Pi_{H[B]}(\lambda)
  &\ \forall\ \lambda\ge t.
  \end{cases}
  \end{aligned}
 \]
 In particular,
 \[
 \begin{aligned}
  \Pi_{H/\mathcal P}(t)
  &=\mathcal S_{H/\mathcal P},\\
  \Pi_{H[B]}(t)
  &=\{B\}
  \qquad \text{for every } B\in\mathcal P.
 \end{aligned}
 \]
\end{enumerate}
\end{lemma}

\begin{proof}
For classical submodular formulations of nesting and
restriction--contraction, see Narayanan~\cite[Theorem~12.3.1,
pp.~636--637, and Section~12.6, pp.~665--668]{Nar97}.
We give a direct proof for the graph objective, including the
boundary values of the parameter.

For part~(1), let $\mathcal A=\Pi_H(x)$ and
$\mathcal B=\Pi_H(y)$. Write $\mathcal A\wedge\mathcal B$
for their common refinement into nonempty part intersections,
and $\mathcal A\vee\mathcal B$ for their finest common coarsening.
The crossing edges of the common refinement are the union
of the two crossing-edge sets, whereas those of the common
coarsening are contained in their intersection. Hence,
\[
 c_H(\delta_H(\mathcal A))+c_H(\delta_H(\mathcal B))
 \ge
 c_H(\delta_H(\mathcal A\wedge\mathcal B))
 +c_H(\delta_H(\mathcal A\vee\mathcal B)).
\]
Also,
\[
 |\mathcal A\wedge\mathcal B|+|\mathcal A\vee\mathcal B|
 \ge |\mathcal A|+|\mathcal B|.
\]
To see this, form a bipartite graph with one vertex for each
part of $\mathcal A$ and of $\mathcal B$, and one edge for each
nonempty intersection. Its edges correspond to the parts of
$\mathcal A\wedge\mathcal B$, and its connected components
correspond to the parts of $\mathcal A\vee\mathcal B$.
A connected component with $k$ vertices has at least $k-1$ edges,
which gives the inequality.
Since $0\le x\le y$ and
$|\mathcal A\wedge\mathcal B|\ge|\mathcal B|$, these inequalities imply
\[
 \Phi_{H,x}(\mathcal A)+\Phi_{H,y}(\mathcal B)
 \ge
 \Phi_{H,x}(\mathcal A\vee\mathcal B)
 +\Phi_{H,y}(\mathcal A\wedge\mathcal B).
\]
Optimality of $\mathcal A$ and $\mathcal B$ gives the reverse
inequality, so equality holds and
$\mathcal A\wedge\mathcal B$ is optimal at $y$.
Because $\mathcal B$ is the finest optimum, this refinement
must equal $\mathcal B$. Thus $\mathcal B$ refines $\mathcal A$.

For part~(2), suppose that a part $B$ induces $k\ge2$
connected components. Replacing $B$ by those components
leaves the crossing capacity unchanged and increases the
number of parts by $k-1$. For $\lambda>0$, this strictly
decreases the objective, contradicting optimality. For
$\lambda=0$, it produces a strictly finer optimum,
contradicting fineness. Thus every part induces a connected graph.
In particular, $\Pi_H(0)$ is the connected-component partition:
it has objective zero, and positive capacities imply that
every zero-cost partition is a coarsening of that partition.

For the contraction identity in part~(3), fix $0\le\lambda\le t$.
By part~(1), $\Pi_H(\lambda)$ coarsens $\mathcal P$.
Lifting gives a bijection between partitions of $H/\mathcal P$
and coarsenings of $\mathcal P$, and for every such partition
$\mathcal Q$,
\[
 \Phi_{H,\lambda}
   \bigl(\operatorname{lift}_{\mathcal P}(\mathcal Q)\bigr)
 =\Phi_{H/\mathcal P,\lambda}(\mathcal Q).
\]
Consequently, the two minimum values coincide. The partition
$\Pi_H(\lambda)$ is available among the lifted partitions
and refines every optimum there. Since lifting preserves
refinement, it is the lift of the finest optimum of
$H/\mathcal P$.

For the restriction identity, fix $\lambda\ge t$.
Part~(1) implies that $\Pi_H(\lambda)$ refines $\mathcal P$.
Every refinement has the form
$\mathcal R=\bigcup_{B\in\mathcal P}\mathcal R_B$, where
$\mathcal R_B$ is a partition of $B$, and its objective satisfies
\[
 \Phi_{H,\lambda}(\mathcal R)
 =\Phi_{H,\lambda}(\mathcal P)
  +\sum_{B\in\mathcal P}\Phi_{H[B],\lambda}(\mathcal R_B).
\]
The first term is fixed, so minimizing over refinements
amounts to minimizing each summand independently.
Since a global finest optimum is among these refinements,
it is the union of the finest optima of the induced graphs.
At $\lambda=t$, the contraction identity lifts to $\mathcal P$,
so the contracted finest optimum is the singleton partition.
The restriction identity at $t$ yields $\mathcal P$ itself,
so the finest optimum within each part $B$ is $\{B\}$.
This proves all the assertions.
\end{proof}

The next lemma shows that an optimum partition separates
the canonical cut hierarchy into the hierarchy between its
parts and the hierarchies within them.

\begin{lemma}
\label{lem:hierarchy-splicing}
Let $H=(V, E, c: E\rightarrow \Z_+)$ be a connected undirected graph.
Let $t\ge0$ and $\Pi=\Pi_H(t)$. The canonical cut hierarchy of $H$
is obtained from that of $H/\Pi$ by replacing each
leaf representing $B\in\Pi$ with the canonical cut
hierarchy of $H[B]$. 
\end{lemma}
\begin{proof}
We first relate the hierarchy of a connected graph $J$
to its finest optimum partitions. If $|V(J)|\ge2$, the
definition of strength implies that $\{V(J)\}$ is the
unique optimum for $0\le\lambda<\beta(J)$. At
$\lambda=\beta(J)$, the nontrivial optimum partitions
are exactly the Tutte--Nash--Williams partitions. Thus
$\Pi_J(\beta(J))$ is the finest such partition and gives
the children of the root.
For every nonsingleton child $B$,
Lemma~\ref{lem:cch-restriction-contraction}(3) gives
\[
 \Pi_{J[B]}(\beta(J))=\{B\}.
\]
This forces $\beta(J[B])>\beta(J)$, because at and above
its strength the finest optimum of a connected graph with
at least two vertices is nontrivial.
Therefore labels strictly increase along every path of
internal nodes from the root towards a leaf.

For any $\lambda\ge0$, the hierarchy recovers
$\Pi_J(\lambda)$ by expanding every internal node whose
label is at most $\lambda$ and retaining the resulting
unexpanded nodes as parts. Indeed, below the root label
the optimum is the one-part partition, and at or above it
Lemma~\ref{lem:cch-restriction-contraction}(3) reduces the
optimum to the induced graphs on the children. Induction
proves the rule. Conversely, these partitions for all
$\lambda$ determine the hierarchy: the first nontrivial
partition determines the root label and its children, and
the same argument applies recursively within each child.

Now $H/\Pi$ is connected, and every $H[B]$ is connected
by Lemma~\ref{lem:cch-restriction-contraction}(2).
By part~(3) of that lemma, the finest optimum of $H/\Pi$
at $t$ is its singleton partition. Thus every internal-node
label in its hierarchy is at most $t$.
For every nonsingleton $B\in\Pi$, the same lemma gives
$\Pi_{H[B]}(t)=\{B\}$, so the root label of its hierarchy,
and hence all its internal-node labels, exceed $t$.
After the stated replacements, the partitions obtained
by the expansion rule are therefore
\[
 \operatorname{lift}_{\Pi}\bigl(\Pi_{H/\Pi}(\lambda)\bigr)
 \qquad\text{for all }0\le\lambda\le t,
\]
and
\[
 \bigcup_{B\in\Pi}\Pi_{H[B]}(\lambda)
 \qquad\text{for all }\lambda\ge t.
\]
Lemma~\ref{lem:cch-restriction-contraction}(3) identifies
both with $\Pi_H(\lambda)$. The hierarchy is consequently
the canonical cut hierarchy of $H$, with all labels preserved.
The assertion also covers a one-vertex contracted graph:
replacing its sole leaf simply returns the hierarchy inserted there.
\end{proof}

The next lemma provides a stopping criterion for our recursive construction. It shows that if the transition from the one-part partition to the singleton partition occurs within a sufficiently narrow parameter interval, then the canonical cut hierarchy consists of a single split.
\begin{lemma}\label{lem:hierarchy-narrow-interval}
Let $H=(V, E, c: E\rightarrow \Z_+)$ be a connected undirected graph with $2\le |V(H)|\le n$, where $n\ge2$.
Suppose that $0\le\ell\le u$ satisfy $\Pi_H(\ell)=\{V(H)\}$,
and $\Pi_H(u)=\mathcal S_H$.
If $u-\ell<1/(n-1)^2$, then the canonical cut hierarchy
of $H$ is a root with one leaf child per vertex, and
its root label is
\[
    \frac{c_H(E(H))}{|V(H)|-1}.
\]
\end{lemma}
\begin{proof}
Every internal node representing a set $S$ is labelled
by $\beta(H[S])$. If its finest Tutte--Nash--Williams
partition has $k$ parts, then this label has the form
$a/(k-1)$ for an integer $a$, with
$1\le k-1\le |S|-1\le n-1$.
Two distinct such labels $a/b$ and $a'/b'$ therefore satisfy
\[
 \left|\frac{a}{b}-\frac{a'}{b'}\right|
 =\frac{|ab'-a'b|}{bb'}
 \ge\frac{1}{(n-1)^2}.
\]

The assumption $\Pi_H(\ell)=\{V(H)\}$ implies
$\ell<\beta(H)$. Labels strictly increase along paths of
internal nodes, as shown in the proof of
Lemma~\ref{lem:hierarchy-splicing}. Also, the expansion
rule proved there and the equality
$\Pi_H(u)=\mathcal S_H$ imply that every internal-node
label is at most $u$.
Thus all internal-node labels lie in $(\ell,u]$.
The width assumption and the preceding separation bound
imply that all these labels are equal.
Strict increase along paths now rules out any internal
node other than the root. The root's children are therefore
the singleton vertices. Since this singleton partition
is a Tutte--Nash--Williams partition of $H$, the root label
is $c_H(E(H))/(|V(H)|-1)$, as claimed.
\end{proof}

\subsection{Algorithm}\label{sec:cch-algo}
Let $G=(V, E, c: E\rightarrow \Z_+)$ be the input graph with $n$ vertices and let $\eta\in (0,1)$ be the desired failure probability. If $n=1$, return its single leaf. We henceforth assume that $n\ge 2$. Initially, find the connected components $C$ of $G$ and call the algorithm to be described below on $G[C]$ with interval $[0, W+1]$ for each component.
If $G$ is connected, return the hierarchy from its only call.
Otherwise, attach the returned hierarchies to a new root labelled zero and return it.
If any recursive call reports failure, the entire algorithm reports failure.
Let 
\begin{align*}
 W &:= \sum_{e\in E(G)} c_e,\\
 \Delta &:= \frac{1}{(n-1)^2}, \\
 L&:=1+\left\lceil\log_2\bigl((W+1)(n-1)^2\bigr)\right\rceil, \text{ and}\\
 \epsilon&:=\frac{\eta}{2nL}.
\end{align*}
We will construct the canonical cut hierarchy recursively and the above parameters do not change across recursive calls. All calls to the algorithm of Theorem~\ref{thm:oracle} use failure probability $\epsilon$ and fresh randomness. Each recursive call takes as input a connected graph $H$ and a parameter interval $[\ell, u]$ for some rational values $0\le\ell\le u$ satisfying 
\begin{equation}\label{eq:hierarchy-invariant}
    \begin{aligned}
        \Pi_H(\ell)&=\{V(H)\},\\
        \Pi_H(u)&=\mathcal S_H.
    \end{aligned}
\end{equation}
For such an input, we describe the recursive construction. 
We first describe the two base cases of the recursion: (1) if $H$ has only one vertex, then we return its only vertex as the leaf. (2) if  $u-\ell <\Delta$, then we return a root labeled by $c_H(E(H))/(|V(H)|-1)$ with one leaf child for each vertex of $H$. 
Next, we describe the recursive cases. 
Let $t$ be the mid-point of the interval. 
We compute the unique finest optimal partition $\Pi$ to the Lagrangian problem on $H$ for $\lambda=t$. Next, we recurse on the graph $H/\Pi$ with the interval $[\ell, t]$ to obtain the canonical cut hierarchy $\mathcal{H}_0$ of $H/\Pi$. Next, for each part $B\in \Pi$, we recurse on the graph $H[B]$ with the interval $[t, u]$ to obtain the canonical cut hierarchy $\mathcal{H}_B$ of the graph $H[B]$ and we replace the leaf representing $B$ in the hierarchy $\mathcal{H}_0$ by the hierarchy $\mathcal{H}_B$. We return the constructed canonical cut hierarchy $\mathcal{H}_0$. See Algorithm \ref{alg:build-hierarchy} for a pseudocode. 

\begin{algorithm}[H]
\caption{\textsc{BuildHierarchy}$(H,\ell,u)$}
\label{alg:build-hierarchy}
\textbf{Input:} A connected graph $H$ and an interval $[\ell,u]$
         satisfying~\eqref{eq:hierarchy-invariant}.\\
\textbf{Output:} The canonical cut hierarchy of $H$, or failure.
\begin{enumerate}
 \item If $|V(H)|=1$, return a leaf representing the only vertex of $H$. 
 \item If $u-\ell <\Delta$, then return a root labelled $c_H(E(H))/(|V(H)|-1)$ with one leaf child for each vertex of $H$. 
 \item $t\leftarrow (\ell+u)/2$. 
 \item Call the algorithm of Theorem~\ref{thm:oracle}, with fresh randomness and failure probability $\epsilon$, to compute $\Pi:=\Pi_H(t)$. If the returned object is not a partition of $V(H)$, report failure. 
 \item Construct $H/\Pi$ and $H[B]$ for every $B\in \Pi$. If some $H[B]$ is disconnected, report failure. 
 \item $\mathcal{H}_0\leftarrow \textsc{BuildHierarchy}(H/\Pi,\ell,t)$
 \item For each $B\in \Pi$,
 \begin{enumerate}
     \item $\mathcal{H}_B\leftarrow \textsc{BuildHierarchy}(H[B],t, u)$.
     \item Replace the leaf representing $B$ in $\mathcal{H}_0$ by the hierarchy $\mathcal{H}_B$.
 \end{enumerate}
 \item Return $\mathcal{H}_0$. 
\end{enumerate}
\end{algorithm}

Each vertex of a recursive graph is stored as an atomic identifier;
the set of original vertices it represents is not expanded.
The returned hierarchy stores a pointer to the leaf corresponding to
each vertex of that recursive graph. Replacing a leaf therefore takes
constant time, apart from computing the hierarchy that replaces it.
If the leaf is the whole hierarchy, replace its root pointer;
no extra node is introduced.

\subsection{Correctness}\label{sec:cch-correctness}
In this section, we prove the correctness of the algorithm. 
\begin{lemma}\label{lem:hierarchy-correct}
Suppose the graph $H$ on input interval $[\ell, u]$ satisfies \eqref{eq:hierarchy-invariant}. 
If each call to the algorithm of Theorem \ref{thm:oracle} returns a finest optimum partition to the Lagrangian problem on $H$ for parameter $t$, 
then Algorithm~\ref{alg:build-hierarchy} terminates, every
recursive call satisfies~\eqref{eq:hierarchy-invariant},
and the algorithm returns the canonical cut
hierarchy of $H$. 
\end{lemma}
\begin{proof}
Consider an input $(H,\ell,u)$ to
Algorithm~\ref{alg:build-hierarchy} satisfying
\eqref{eq:hierarchy-invariant}, and suppose that all calls
to the algorithm of Theorem~\ref{thm:oracle} return the
finest optimum partition for their respective input graph
and parameter. We first show that the invariant is preserved.
In a nonterminal call, the partition returned at
$t=(\ell+u)/2$ is $\Pi=\Pi_H(t)$.
The contraction identity in
Lemma~\ref{lem:cch-restriction-contraction}(3), applied
at $\ell$ and $t$, gives
\[
\begin{aligned}
 \Pi_{H/\Pi}(\ell)&=\{V(H/\Pi)\},\\
 \Pi_{H/\Pi}(t)&=\mathcal S_{H/\Pi}.
\end{aligned}
\]
The restriction identity in the same lemma gives,
for every $B\in\Pi$,
\[
\begin{aligned}
 \Pi_{H[B]}(t)&=\{B\},\\
 \Pi_{H[B]}(u)&=\mathcal S_{H[B]}.
\end{aligned}
\]
For the last equality, the union of the finest optima of
the induced graphs equals $\Pi_H(u)=\mathcal S_H$, so each
of those optima must be its singleton partition.
Thus every child call satisfies the invariant.
The contracted graph $H/\Pi$ is connected because $H$ is
connected, and every $H[B]$ is connected by
Lemma~\ref{lem:cch-restriction-contraction}(2).
The returned partition is valid, so neither check in the
algorithm reports failure. Induction proves that every
recursive input is valid and satisfies the invariant.

Next, we prove termination. Write $w=u-\ell$ for the
width of the interval in the initial call under consideration.
If $w<\Delta$, this call is a base case.
Otherwise, an interval at depth $d$ below this call has
width $w/2^d$. Since $\Delta>0$, this width becomes less
than $\Delta$ after finitely many halvings.
Every branch therefore has bounded depth. Each call has
finitely many children, so the entire recursion terminates.

Finally, we prove correctness by induction on the height
of the recursion below a call.
If $|V(H)|=1$, the returned leaf is the canonical cut
hierarchy of $H$.
If $|V(H)|\ge2$ and $u-\ell<\Delta$, the invariant and
Lemma~\ref{lem:hierarchy-narrow-interval} show that the
returned star, including its root label, is correct.
In the remaining case, the induction hypothesis implies
that the recursive calls return the canonical cut
hierarchies of $H/\Pi$ and of $H[B]$ for every $B\in\Pi$.
By Lemma~\ref{lem:hierarchy-splicing}, replacing each leaf
representing $B$ in the former hierarchy by the latter
hierarchy gives the canonical cut hierarchy of $H$.
This is exactly the hierarchy returned by
Algorithm~\ref{alg:build-hierarchy}.
\end{proof}

\begin{corollary}
Suppose that every call to the algorithm of
Theorem~\ref{thm:oracle} returns the finest optimum partition
for its input graph and parameter.
Then the algorithm described in Section~\ref{sec:cch-algo}
returns the canonical cut hierarchy of $G$.
\end{corollary}
\begin{proof}
If $n=1$, the algorithm returns the only vertex as a
leaf, which is the required hierarchy. Assume henceforth
that $n\ge2$.
Let $C$ be a connected component of $G$, and put $H=G[C]$.
We verify that the initial call
$\textsc{BuildHierarchy}(H,0,W+1)$ satisfies
\eqref{eq:hierarchy-invariant}.
Since $H$ is connected and its capacities are positive,
$\Pi_H(0)=\{V(H)\}$.
If $|V(H)|=1$, its only partition also equals
$\mathcal S_H$, so the other equality is immediate.
Otherwise, let $r=|V(H)|$ and consider any partition
$\mathcal Q\ne\mathcal S_H$ of $V(H)$.
Since $r-|\mathcal Q|\ge1$ and $c_H(E(H))\le W$, we have
\[
\begin{aligned}
 \Phi_{H,W+1}(\mathcal S_H)-\Phi_{H,W+1}(\mathcal Q)
 &=c_H(E(H)\setminus\delta_H(\mathcal Q))
   -(W+1)(r-|\mathcal Q|)\\
 &\le W-(W+1)\\
 &<0.
\end{aligned}
\]
Consequently, $\mathcal S_H$ is the unique optimum at
$W+1$, proving the invariant for this initial call.

By the hypothesis and Lemma~\ref{lem:hierarchy-correct},
the call on each component terminates and returns its
canonical cut hierarchy.
If $G$ is connected, the algorithm returns this hierarchy.
If $G$ is disconnected, it attaches the roots of the
component hierarchies to a new root labelled zero.
By the definition of the canonical cut hierarchy of a
disconnected graph, the resulting hierarchy is exactly
the canonical cut hierarchy of $G$.
\end{proof}

\subsection{Runtime}\label{sec:cch-runtime}
In this section, we bound the runtime of the algorithm and prove Theorem \ref{thm:cch}. Let $N=m+n$, where $m$ is the number of edges and $n$ is the number of vertices in the input graph $G$. For a recursive input graph $H$, let $n_H:=|V(H)|$ and $m_H:=|E(H)|$. The following lemma bounds the total graph size and the number of calls to the algorithm of Theorem~\ref{thm:oracle} at a fixed recursion depth. 
\begin{lemma}\label{lem:hierarchy-size-accounting}
At any fixed recursion depth, let $\mathcal A$ be the collection
of recursive calls whose graphs have at least two vertices.
Then
\begin{align*}
    \sum_{H\in\mathcal A}m_H&\le m,\\
    \sum_{H\in\mathcal A}(n_H-1)&\le n-1,\text{ and}\\
    \sum_{H\in\mathcal A}n_H&\le2(n-1).
\end{align*}
In particular, there are at most $n-1$ such calls per depth
and at most $(n-1)L$ calls in total to the algorithm of Theorem \ref{thm:oracle}. These bounds hold even if some of the returned partitions are not optimal.
\end{lemma}
\begin{proof}
Consider a nonterminal call on $H$ that returns a
partition $\Pi$ with $k$ parts. Each graph edge belongs
to exactly one child graph: a crossing edge belongs to
$H/\Pi$, and an edge within a part $B$ belongs to $H[B]$.
Because parallel edges are retained, we have
\[
 m_{H/\Pi}+\sum_{B\in\Pi}m_{H[B]}=m_H.
\]
The corresponding identity for vertices is
\[
 (k-1)+\sum_{B\in\Pi}(|B|-1)=n_H-1.
\]
For the initial connected-component calls, the edge counts
sum to $m$, and the quantities $n_H-1$ sum to at most $n-1$.
Discarding singleton graphs, terminal calls, or calls after
a reported failure can only decrease these sums.
Induction on the depth proves the first two inequalities.
Every graph in $\mathcal A$ contributes at least one to
the second sum, so $|\mathcal A|\le n-1$, and
\[
 \sum_{H\in\mathcal A}n_H
 =\sum_{H\in\mathcal A}(n_H-1)+|\mathcal A|
 \le2(n-1).
\]
Independently of which partitions are returned, every
recursive interval is obtained by halving its parent's
interval. Only depths $0,\ldots,L-1$ can contain calls
to the algorithm of Theorem~\ref{thm:oracle}.
There are at most $n-1$ such calls at each depth, proving
the total bound. None of these arguments requires
optimality of the returned partitions.
\end{proof}

We now prove Theorem \ref{thm:cch}. 
\begin{proof}[Proof of Theorem~\ref{thm:cch}]
Run the algorithm of Section~\ref{sec:cch-algo}.
For $n=1$, it returns the only leaf in constant time;
assume henceforth that $n\ge2$.
Conditional on the history preceding any call to the
algorithm of Theorem~\ref{thm:oracle}, fresh randomness
gives failure probability at most $\epsilon$.
Lemma~\ref{lem:hierarchy-size-accounting} bounds the number
of such calls, on every execution, by $(n-1)L$.
A union bound on the first incorrect returned partition gives
\[
 \Pr[\text{some call returns an incorrect partition}]
 \le(n-1)L\epsilon<\eta.
\]
On the complementary event,
Lemma~\ref{lem:hierarchy-correct} proves that the returned
hierarchy is correct.

We next bound the additional arithmetic operations.
Set $\Lambda:=\log(nL/\eta)$.
For a call on $H$ with $n_H\ge2$, Theorem~\ref{thm:oracle}(1)
uses $O(\Lambda)$ maximum-flow computations and
\[
 O\!\left((m_H+n_H)\log^2 n\log N\,\Lambda\right)
\]
additional expected arithmetic operations, since
$\log(n_H/\epsilon)=O(\Lambda)$.
Lemma~\ref{lem:hierarchy-size-accounting} bounds
$\sum_H(m_H+n_H)$ by $O(N)$ at each depth.
Summing over the $L$ depths that can contain these calls gives
\[
 O\!\left(N\log^2 n\log N\,L\Lambda\right).
\]
Constructing the contracted and induced graphs requires
one scan of the vertices and edges of each parent graph.
The partition and connectivity checks, evaluation of the base-case
labels, and leaf replacements also take $O(m_H+n_H)$
operations per nontrivial call. Creating and returning
singleton calls is charged to their parent's vertex count;
initial singleton components take $O(n)$ time in total.
All this work is $O(N(L+1))$, including the terminal depth,
and is absorbed by the claimed bound.
The output has $n$ leaves and at least two children at every
internal node, so it has at most $2n-1$ nodes.

To obtain the claimed number of maximum-flow computations,
process the recursion one depth at a time.
At a fixed depth, advance each call to the algorithm of
Theorem~\ref{thm:oracle} until its next required maximum-flow
computation. Form the disjoint union of the resulting
networks, identifying their sources with a common source
and their sinks with a common sink.
The maximum-flow value in this combined network is the
sum of the individual maximum-flow values: feasible flows
combine, and a flow in the combined network restricts to
a feasible flow in each constituent network.
Consequently, an integral maximum flow in the combined
network restricts to an integral maximum flow in every
constituent network. Return these restrictions to the
corresponding computations and continue in this way.
Each call requires at most $O(\Lambda)$ maximum-flow
computations, so there are at most $O(\Lambda)$ combined
computations per depth, and $O(L\Lambda)$ in total.
Theorem~\ref{thm:oracle}(1) and
Lemma~\ref{lem:hierarchy-size-accounting} imply that every
combined network has $O(N)$ vertices and $O(N\log^2 n)$ arcs.
Constructing these networks and distributing the resulting
flows take linear work in their sizes, already covered
by the additional-operation bound. A completed call
simply does not participate in subsequent computations.
Processing calls by depth changes only their order;
the recursive outputs are combined as in Algorithm~\ref{alg:build-hierarchy}.

We now bound the capacities using only
Theorem~\ref{thm:oracle}(1).
Every midpoint queried at a depth less than $L$ has a
representation $t=p/d$ with positive integers satisfying
$d\le2^L$, and $p\le(W+1)2^L$.
Moreover, $2^L\le4(W+1)(n-1)^2$. 
Every edge of a recursive graph retains an original
capacity, so its maximum capacity $U_H$ is at most $W$.
Each call with at least two vertices has a connected graph,
and hence has an edge; thus $U_H$ is well-defined.
Theorem~\ref{thm:oracle}(1) supplies integer-capacity
networks with capacities bounded by
\[
 \bigl(n_H(U_H+1)(p+d+1)\bigr)^{O(1)}
 \le \bigl(n(W+1)\bigr)^{O(1)}.
\]
Identifying sources and sinks does not change any capacity;
we retain parallel arcs separately. This proves the
integer-capacity guarantee for the combined networks.

Finally, suppose the original capacities are polynomially
bounded in $N=m+n$ and $\eta=n^{-a}$ for a fixed $a>0$.
Then $W$ is polynomially bounded in $N$,
$L=O(\log N)$, and $\Lambda=O(\log N)$.
The preceding parameter bounds show that $U_H,p,d$ are
polynomially bounded in $N$ for every recursive call.
Also,
\[
 \frac{1}{\epsilon}=2nL\,n^a
\]
is polynomially bounded in $N$.
Apply Theorem~\ref{thm:oracle}(2) with the common size
parameter $S=N$ and a fixed exponent large enough for all
these bounds. Since we supply integral maximum flows,
the additional expected time for a call on $H$ is
\[
 O\!\left((m_H+n_H)\log^{O(1)}N\right).
\]
The size accounting over $L=O(\log N)$ depths therefore
gives $\widetilde O(N)$ additional expected time.
The construction and combination steps also take
$\widetilde O(N)$ time, since their capacities, parameters,
and node labels have $O(\log N)$-bit encodings.

Every combined maximum-flow network has
$\widetilde O(N)$ size and integer capacities bounded
by a fixed polynomial in $N$.
If needed, add $N$ isolated unit-capacity arcs, unconnected
to the source and sink. This preserves the maximum-flow
problem and makes its size at least $N$.
The deterministic maximum-flow algorithm
of~\cite{BCKLPPSS23} then computes an integral maximum
flow in $N^{1+o(1)}$ time.
There are $O(L\Lambda)=O(\log^2 N)$ such computations,
so their total time is $N^{1+o(1)}$.
Combining this with the additional expected time proves
the claimed bound $(m+n)^{1+o(1)}$.
\end{proof}

\paragraph{Acknowledgements.} We thank Chandra Chekuri for insighful discussions and providing feedback that helped improve the manuscript. 
For AI disclosure, see Section \ref{sec:techniques}. 

\bibliographystyle{abbrv}
\bibliography{references}

\end{document}